\documentclass[twocolumn]{aastex631}

\usepackage{xspace}

\newcommand{\source}{4U~1636--536\xspace}
\newcommand{\nicer}{NICER}
\newcommand{\nustar}{NuSTAR}
\newcommand{\fluxcgs}{erg~s$^{-1}$~cm$^{-2}$}
\newcommand{\flencecgs}{erg~cm$^{-2}$}
\newcommand{\fa}{$f_a$~}
\extrafloats{100}
\shorttitle{X-ray Bursts from \source}
\shortauthors{G\"uver et al.}
\graphicspath{{./}}

\begin{document}

\title{Burst - Disk Interaction in \source as observed by \nicer}

\author[0000-0002-3531-9842]{Tolga G\"uver}
\affiliation{Istanbul University, Science Faculty, Department of Astronomy and Space Sciences, Beyaz\i t, 34119, Istanbul, Turkey}
\affiliation{Istanbul University Observatory Research and Application Center, Istanbul University 34119, Istanbul Turkey}

\author[0000-0002-5665-3452]{Z. Funda Bostanc\i~}
\affiliation{Istanbul University, Science Faculty, Department of Astronomy and Space Sciences, Beyaz\i t, 34119, Istanbul, Turkey}
\affiliation{Istanbul University Observatory Research and Application Center, Istanbul University 34119, Istanbul Turkey}

\author[0000-0002-4729-1592]{Tu\u{g}ba Boztepe}
\affiliation{Istanbul University, Graduate School of Sciences, Department of Astronomy and Space Sciences, Beyaz\i t, 34119, Istanbul, Turkey}

\author[0000-0002-5274-6790]{Ersin G\"o\u{g}\"u\c{s}}
\affiliation{Faculty of Engineering and Natural Sciences, Sabanc\i~University, Orhanl\i-Tuzla 34956, \.Istanbul, Turkey }

\author[0000-0002-7252-0991]{Peter Bult}
\affiliation{Department of Astronomy, University of Maryland, College Park, MD 20742, USA}
\affiliation{Astrophysics Science Division, NASA Goddard Space Flight Center, Greenbelt, MD 20771, USA}

\author[0000-0001-6530-0221]{Unnati Kashyap}
\affiliation{Department of Astronomy, Astrophysics and Space Engineering (DAASE),
Indian Institute of Technology Indore, Khandwa Road, Simrol, Indore 453552, India}

\author[0000-0002-9736-9538]{Manoneeta Chakraborty}
\affiliation{Department of Astronomy, Astrophysics and Space Engineering (DAASE),
Indian Institute of Technology Indore, Khandwa Road, Simrol, Indore 453552, India}

\author[0000-0001-8128-6976]{David R. Ballantyne}
\affiliation{Center for Relativistic Astrophysics, School of Physics, Georgia Institute of Technology, 837 State Street, Atlanta, GA 30332-0430, USA}

\author[0000-0002-8961-939X]{R.~M.~Ludlam}\thanks{NASA -- Einstein Fellow}
\affiliation{Cahill Center for Astronomy and Astrophysics, California Institute of Technology, Pasadena, CA 91125, USA}

\author[0000-0002-0380-0041]{C.~Malacaria}
\affiliation{Universities Space Research Association, Science and Technology Institute, 320 Sparkman Drive, Huntsville, AL 35805, USA}

\author[0000-0002-6789-2723]{Gaurava K. Jaisawal}
\affiliation{National Space Institute, Technical University of Denmark, Elektrovej 327-328, DK-2800 Lyngby, Denmark}

\author[0000-0001-7681-5845]{Tod E. Strohmayer}
\affiliation{Astrophysics Science Division and Joint Space-Science Institute, NASA's Goddard Space Flight Center, Greenbelt, MD 20771, USA}

\author[0000-0002-6449-106X]{Sebastien~Guillot}
\affil{Institut de Recherche en Astrophysique et Plan\'{e}tologie, UPS-OMP, CNRS, CNES, 9 avenue du Colonel Roche, BP 44346, F-31028 Toulouse Cedex 4, France}

\author[0000-0002-0940-6563]{Mason Ng}
\affiliation{MIT Kavli Institute for Astrophysics and Space Research, Massachusetts Institute of Technology, Cambridge, MA 02139, USA}



\begin{abstract}

We present the detection of 51 thermonuclear X-ray bursts observed from \source by the Neutron Star Interior Composition Explorer~(\nicer)~over the course of a three year monitoring campaign. We performed time resolved spectroscopy for 40 of these bursts and showed the existence of a strong soft excess in all the burst spectra. The excess emission can be characterized by the use of a scaling factor (\fa method) to the persistent emission of the source, which is attributed to the increased mass accretion rate on to the neutron star due to Poynting-Robertson drag. The soft excess emission can also be characterized by the use of a model taking into account the reflection of the burst emission off of the accretion disk. We also present time resolved spectral analysis of 5 X-ray bursts simultaneously observed by \nicer~and AstroSat, which confirm the main results with even  greater precision. Finally, we present evidence for Compton cooling using 7 X-ray bursts observed contemporaneously with \nustar, by means of a correlated decrease in the hard X-ray lightcurve of \source as the bursts start. 

\end{abstract}
\keywords{X-rays: bursts --- accretion disks --- \source ---}


\section{Introduction} \label{sec:intro}

X-ray bursts observed from low mass X-ray binaries have long been attributed to thermonuclear burning of accreted material onto the neutron star \citep[see][]{1975ApJ...195..735H,1978ApJ...220..291L,1993SSRv...62..223L,2006csxs.book..113S}. Observations of these events have been used to infer the physical properties of neutron stars \citep[see][]{2016ApJ...820...28O} as well as the accretion environment around these objects \citep{2015MNRAS.454.1371W,2017JApA...38...39P,2018SSRv..214...15D}. In particular, observations of bursts showing evidence for photospheric radius expansion allowed the measurement of the Eddington limit for the neutron stars in these systems \citep[see][]{1978Natur.274..650V,1987A&A...172L..20V,1990A&A...237..103D,2008MNRAS.387..268G,2012ApJ...747...77G}, which are further used to constrain the masses and radii of neutron stars \citep{2006Natur.441.1115O,2009ApJ...693.1775O,2010ApJ...712..964G,2010ApJ...719.1807G,2010PhRvD..82j1301O,2012ApJ...747...77G,2014ApJ...784..123L,2016ApJ...820...28O,2016EPJA...52...18S,2018MNRAS.476..421S}. Although a variety of possible systematic uncertainties in these measurements have been studied using the available data \citep[see][]{2012ApJ...747...77G,2012ApJ...747...76G,2014MNRAS.445.4218K,2016ApJ...829...48G}, complications that may be involved in the physics of the bursts and accretion geometry have remained open questions \citep{2017A&A...608A..31N,2017MNRAS.472...78K}. 

A significant fraction of research related to thermonuclear X-ray bursts has been performed by the Rossi X-ray timing Explorer (RXTE), Proportional Counter Array (PCA). The energy range, where RXTE/PCA had maximum sensitivity (2.5--25.0~keV) matched very well with the typical temperatures of the thermal emission arising from thermonuclear X-ray bursts, which allowed for detailed observations of these events \citep[see][]{Galloway2008,2020ApJS..249...32G}. However, apart from a few exceptions \citep[][]{2004ApJ...602L.105B,2013ApJ...772...94W,2015ApJ...801...60W}, the effects of bursts on the accretion disk and surrounding material remained mostly unstudied due to the lack of low energy sensitivity of RXTE/PCA. Unlike RXTE, the Neutron Star Interior Composition Explorer (NICER) allows for a systematic study of thermonuclear X-ray bursts in the soft X-rays and allows to probe the effects of bursts on the surrounding material \citep[see][]{2018ApJ...855L...4K,2018ApJ...856L..37K,2019ApJ...885L...1B,2020MNRAS.499..793B,2022MNRAS.510.1577G}. Recently, we studied all the X-ray bursts from Aql X-1 prior to the end of 2021 using \nicer~ observations and showed that in approximately 68\% of the spectra there is statistical evidence for an excess emission which can be characterized by the application of a scaling factor to the persistent emission of the system \citep{2022MNRAS.510.1577G}. Supporting earlier findings, this indicates a significant increase in pre-burst emission especially around the peak flux moments of the bursts. This increase is attributed to the increased mass accretion rate on to the neutron star due to Poynting-Robertson drag caused by the burst emission and supported by simulations of accretion disks illuminated by thermonuclear X-ray bursts \citep{2018ApJ...867L..28F,2020NatAs...4..541F}.  
\cite{2022MNRAS.510.1577G} showed that the X-ray spectra can also be fit with models taking into account reflection of the burst emission off of the accretion disk.

We here continue on the same effort using \nicer~ observations of the low mass X-ray binary \source which frequently shows X-ray bursts. \source was first observed by Uhuru and Copernicus \citep{1972ApJ...178..281G, 1974MNRAS.169....7W} and the first X-ray burst was detected by the 8th Orbiting Solar Observatory~(OSO-8) \citep{1976IAUC.3000U...1S}. It has been one of few low mass X-ray binaries that frequently displays X-ray bursts. Within the MINBAR catalog \citep{2020ApJS..249...32G}, the mean burst rate is given as 0.26~hr$^{-1}$, which makes it the fourth most frequent X-ray burster after IGR~J17480--2446, GS~1826--24, and 4U~1728--34. A total of 664 bursts from \source have been cataloged in MINBAR, with 140 of those bursts shoving evidence for photospheric radius expansion \citep{2020ApJS..249...32G}. From \source, various types of X-ray bursts such as double peaked X-ray bursts \citep{2006ApJ...636L.121B}, triple peaked X-ray bursts \citep{1986MNRAS.221..617V,2009MNRAS.398..368Z,2019MNRAS.482.4397B}, short recurrence time bursts \citep{2010ApJ...718..292K} have been reported.

The spin frequency of the neutron star as inferred from burst oscillations and the orbital period of the system are $\sim$ 581~Hz \citep{1997IAUC.6541....1Z,1998ApJ...498L.135S,2002ApJ...577..337S} and 3.8~hr \citep{1998ApJ...503L.147S}, respectively. The distance of \source was calculated to be 4.4$_{-3.1}^{+1.6}$~kpc based on Gaia DR2 results \cite{2021MNRAS.502.5455A}. The orbital inclination angle of \source was limited to $36^{\circ}$ -- $74^{\circ}$ by \cite{2006MNRAS.373.1235C}.
Moreover, \cite{2008ApJ...688.1288P} has set a lower limit of $64^{\circ}$ on the disk inclination using the high signal-to-noise ratio of source spectra.
\source is known to show a broad, asymmetrical iron emission line covering the energy range of 4--9~keV that is consistent with the relatively extended Fe $K\alpha$ emission from the inner accretion disk \citep{2008ApJ...688.1288P, 2010ApJ...720..205C,2013MNRAS.432.1144S,2014MNRAS.440.1165L,2017ApJ...836..140L,2021MNRAS.504.1331M}.

\source has been one of the first X-ray binaries where evidence for deviations from a pure blackbody emission during the X-ray bursts have been reported \citep{2013ApJ...772...94W}. More recently, \cite{2022MNRAS.509.3989K} reported the detection of such excess near the peaks of the bursts in broadband AstroSat observations. They fit the X-ray spectra extracted from LAXPC and SXT with either two blackbody functions with a constant background assumption or with a model that includes a scaling factor for the pre-burst emission and another blackbody model for the burst emission. Although the data presented in \cite{2022MNRAS.509.3989K} were not sensitive enough to differentiate among these models, they statistically demonstrated the requirement of the existence of a second component to describe the burst spectra near the peaks of the bursts. As with other examples the excess can be interpreted as the re-emission/reprocessing of the photons by the accretion disk/corona or scattering of the photons in the neutron star atmosphere or the enhanced persistent emission due to the Poynting–Robertson drag \citep{2013A&A...553A..83I, 2013ApJ...767L..37D,2013ApJ...772...94W,2017ApJ...836..111K,2018ApJ...855L...4K}. 

We here report the detection and time resolved spectral and temporal analysis of the bursts observed from \source by \nicer~since June 2017. In total we detected 51 X-ray bursts. We note that some of the bursts reported here have already been presented by \cite{2022A&A...660A..31Z}. Although our results are mostly in agreement, we include those bursts as well for completeness. We find that six of these bursts have been observed simultaneously with \nustar~and also another five bursts have also been simultaneously observed with AstroSat. We also present the results of time resolved spectral analysis of the bursts observed simultaneously with AstroSat and \nicer. Finally we use some of the bursts detected with \nustar~ to search for evidence for Compton cooling in the system.

\section{Observation and Data Analysis} 
\label{sec:analysis}

Since the start of its operations \nicer~performed 138 different observations of \source (covering the OBSIDs 0050080104 to 0050080113, 1050080101 to 1050080207, 2050080201 to 2050080225, and 3050080201 to 3050080206) from June 26th 2017 to March 30th 2020. 
These observations led to a total of 518.58~ks of unfiltered exposure time. We analyzed this data using HEASOFT v6.29c \citep{2014ascl.soft08004N}, NICERDAS v8c and the calibration files distributed as of July 07 2020. Clean event files were generated using \texttt{nicerl2} and standard filtering criteria\footnote{https://heasarc.gsfc.nasa.gov/lheasoft/ftools/headas/nimaketime.html}, which resulted in a total of 297.559~ks clean exposure time. We barycentered the time of arrivals of the events using the source coordinates~(J2000) R.A. 16h40m55.51s DEC. -53$^{\circ} 45^{\prime}05.0 ^{\prime\prime}$ and JPLEPH.430 ephemerides \citep{2014IPNPR.196C...1F}.

\begin{figure}
    \centering
    \includegraphics[scale=0.5]{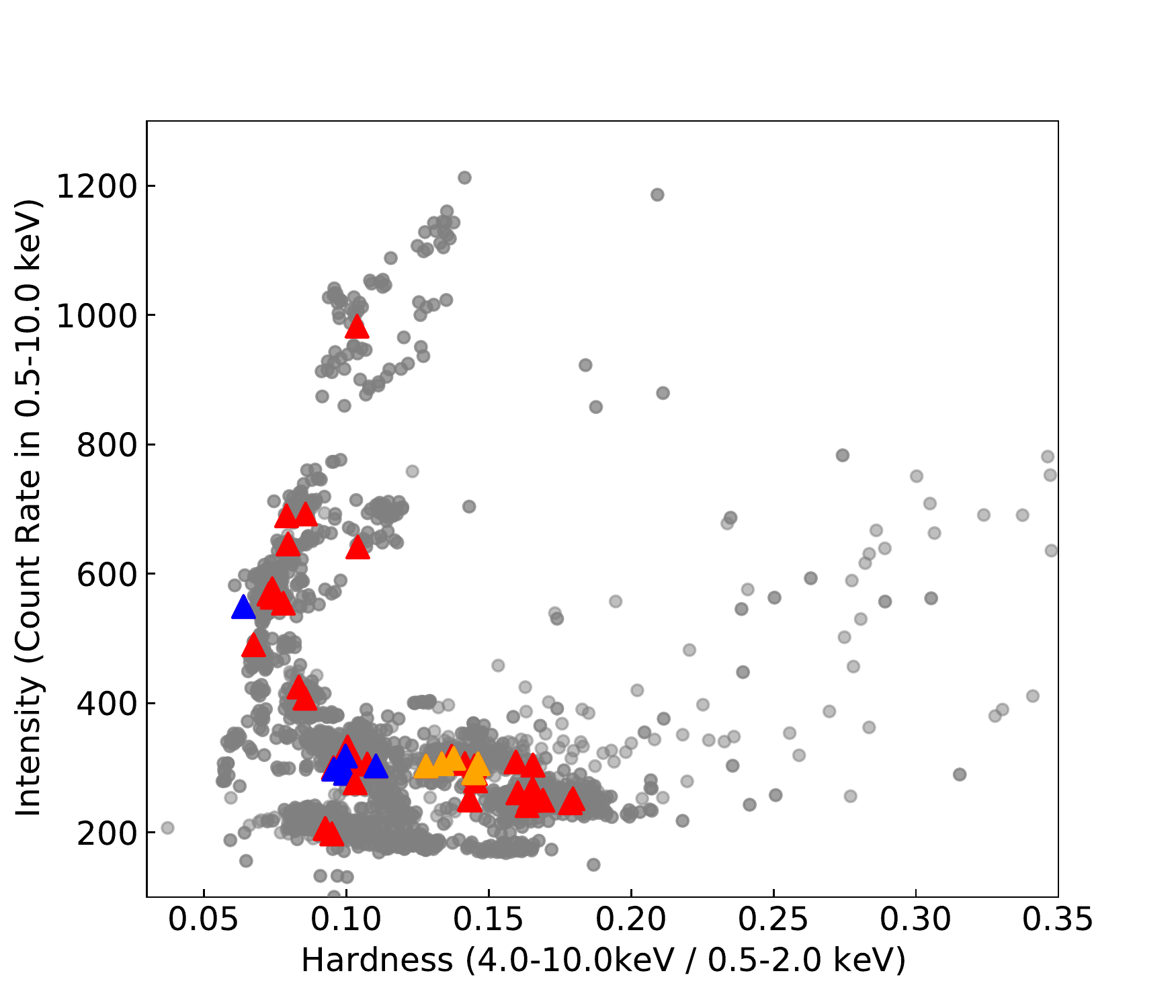}
    \caption{ Hardness-Intensity diagram from all NICER observations of \source. Hardness ratio is defined as the ratio of the count rate in 4.0--10.0~keV and the 0.5--2.0~keV bands. The locations of the detected X-ray bursts are indicated by red triangles. We also show in blue and orange the locations of X-ray bursts simultaneously detected with AstroSat and \nustar, respectively. 
    }
    \label{fig:hid}
\end{figure}

We first generated a hardness intensity diagram using all of the \nicer~ observations to see if there is any state dependence on burst occurrence. For this purpose, we extracted lightcurves using only clean event files with a time resolution of 128~s in various energy bands as defined in \cite{2021ApJ...907...79B}. We used the 0.5--10.0~keV band for the total intensity and calculated the hardness using the ratio of observed count-rates in 4.0--10.0 and 0.5--2.0~keV bands. The resulting hardness intensity diagram is shown in \autoref{fig:hid} together with the spectral state of the source just prior to the observed X-ray bursts. 

We performed our search for bursts using all of the unfiltered events. In each case, we searched for a 4$\sigma$ deviation from the 25~s long background level in the 0.5--10.0~keV lightcurves having 0.25~s time resolutions. We then visually inspected all the candidates for the well known fast rise exponential decay profile. In total we identified 51 events as thermonuclear X-ray bursts. Note that 33 of these bursts can be seen in clean events while for 18 bursts we needed unfiltered data. Some of the basic properties of these bursts are given in \autoref{tab:burst_prop} and lightcurves of some of the bursts are given in the Appendix. In \autoref{tab:burst_prop} burst start times are defined as the first moment the count rate increases by 4$\sigma$ above the average count rate of the last 100~s. The rise time is defined as the first moment the count rate reaches 98\% of the maximum recorded rate during the burst. Finally the e-folding time is defined as the time the count rate decreased by a factor of $e$ after the peak moment.

During our analysis we found out that there are AstroSat (OBSID 9000001574 and 9000002316, starting on 2017 February 10 and 2018 August 18, respectively) and \nustar~(OBSID 30401014002, starting on 2019 April 27) observations partly simultaneous with some of the \nicer~observations. We also include here some of the bursts observed during these observations. For details of the \nustar~observation we refer to \cite{2022arXiv220207379R}. For the analysis of the \nustar~data we again used HEASOFT v6.29c, specifically the \texttt{nupipeline} and \texttt{nuproducts} with the latest available calibration files as of October 7th 2021. The total exposure time of the \nustar~observation is 92 ks. Within this observation we identified 15 X-ray bursts. After applying the barycentric correction we find that 6 thermonuclear X-ray bursts were observed simultaneously with \nicer. These correspond to bursts 42, 43, 44, 45, 46, 50 in \autoref{tab:burst_prop}. 
In addition to these bursts we also identified that there are two additional bursts in the unfiltered event file of \nustar~but filtered out when generating the clean event file, which were also simultaneously observed with \nicer. The bursts that are only in the unfiltered event files of \nustar~are 48th and 49th bursts in our \nicer~burst catalog. 

We also searched the AstroSat/LAXPC observations of 4U 1636–536 between 2017 June 21 and 2018 August 2 and detected 27 bursts. We use event analysis (EA) mode data, where each photon's arrival times and energies are recorded. The total exposure times of the LAXPC observations are 19.041 ks and 210.982 ks, respectively. To analyse the LAXPC data, process the light curves and spectra, and extract background and response files, \texttt{LaxpcSoft} software v3.0 is used \citep{2017ApJS..231...10A}. Due to low gain, we do not use LAXPC10 data, and LAXPC30 data are not useful due to gain instability caused by the gas leakage. We use data from LAXPC20 only, as it provides the maximum gain and sensitivity. After the barycentric correction, we identified 5 bursts that were simultaneously observed with AstroSat and NICER, these bursts are
burst 7, 22, 23, 27, and 28.

\startlongtable
\begin{deluxetable*}{ccccccc}
    \tablehead{ \colhead{BID} & \colhead{MJD} & \colhead{OBSID} & \colhead{Peak Rate\tablenotemark{a}}  & \colhead{Pre-burst Rate} & \colhead{Rise Time} &  \colhead{e-folding Time} \\
    & (TDB) &  & \colhead{(counts~s$^{-1}$)} & \colhead{(counts~s$^{-1}$)} & \colhead{(s)}& \colhead{(s)} }
        
    \tablecaption{Some basic parameters of the detected bursts.\label{tab:burst_prop}}
    \startdata
    $1^\diamond$& 57959.158528 & 1050080103 & 1448 $\pm$ 59 & 333.0 $\pm$ 3.0 & 8.5 & 28.5 \\
    2 & 57975.740335 & 1050080112 & 2568 $\pm$ 84 & 980.8 $\pm$ 3.6 & 2.0 & 9.0 \\
    3 & 57982.091498 & 1050080119 & 1890 $\pm$ 70 & 576.8 $\pm$ 2.3 & 6.3 & --\tablenotemark{b} \\
   $4^\diamond$& 57982.994681 & 1050080119 & 2039 $\pm$ 73  & 645.8 $\pm$ 2.6 & 3.8 & 12.0 \\
    5 & 57989.366494 & 1050080126 & 5140 $\pm$ 107 & 651.3 $\pm$ 2.7 & 1.5 & 8.5 \\
    6 & 58028.477201 & 1050080128 & 7741 $\pm$ 129 & 590.0 $\pm$ 2.7 & 2.0 & 3.0 \\
    7$^*$ & 58029.255800 & 1050080129 & 622 $\pm$ 49   & 591.2 $\pm$ 2.2 & 3.0 &  12.0 \\
    $8^\diamond$& 58059.770862 & 1050080133 & 1514 $\pm$ 63  & 492.6 $\pm$ 2.7 & 3.8 & 7.5 \\
    9 & 58131.664792 & 1050080136 & 1953 $\pm$ 66  & 243.8 $\pm$ 1.6 & 5.3 & 15.0 \\
    $10^\diamond$ & 58131.669711 & 1050080136 & 661 $\pm$ 42  & 250.3 $\pm$ 1.6 & 3.3 & 10.5 \\
    $11^\diamond$ & 58132.480698 & 1050080137 & 2041 $\pm$ 65 & 106.4 $\pm$ 15.1 & 2.5 & 21.0 \\
    $12^\diamond$ & 58132.490322 & 1050080137 & 1678 $\pm$ 61 & 226.8 $\pm$ 1.6 & 4.5 & 10.5 \\
    13 & 58132.499211 & 1050080137 & 620 $\pm$ 40  & 216.7 $\pm$ 1.6 & 3.0  &  8.5 \\
    14 & 58150.624493 & 1050080143 & 1977 $\pm$ 67 & 316.3 $\pm$ 17.7 & 5.8 & 23.0 \\
    $15^\diamond$& 58236.932412 & 1050080149 & 8715 $\pm$ 138 & 805.9 $\pm$ 2.7 & 1.5 & 2.0 \\
    $16^\diamond$& 58237.064383 & 1050080150 & 2207 $\pm$ 76 & 690.0 $\pm$ 2.7 & 3.75 & 11.0 \\
    $17^\diamond$& 58237.829134 & 1050080150 & 2026 $\pm$ 63 & 10.3 $\pm$ 0.1 & 2.8 & 70.5 \\
    18 & 58238.417832 & 1050080151 & 1727 $\pm$ 66 & 492.2 $\pm$ 2.7 & 4.8 & 12.5 \\
    19 & 58308.311685 & 1050080158 & 1589 $\pm$ 62 & 376.4 $\pm$ 2.6 & 4.8 & 22.0 \\
    20 & 58312.507264 & 1050080160 & 1849 $\pm$ 68 & 495.6 $\pm$ 2.6 & 3.5 & 21.5 \\
    21 & 58339.052711 & 1050080173 & 1855 $\pm$ 64 & 2559 $\pm$ 6.5 & 7.0 & 19.0 \\
    22$^*$& 58348.422270 & 1050080175 & 1708 $\pm$ 63 & 311.1 $\pm$ 2.2 & 6.8 & 16.0 \\
    23$^*$& 58348.562250 & 1050080175 & 1685 $\pm$ 63 & 308.3 $\pm$ 2.0 & 5.5 & 24.0 \\
    24 & 58348.677453 & 1050080175 & 2076 $\pm$ 69 & 344.6 $\pm$ 2.2 & 4.0 & 15.5 \\
    25 & 58348.803743 & 1050080175 & 599 $\pm$ 41  & 260.0 $\pm$ 11.1 & 2.5 & 14.5 \\
    26 & 58348.934944 & 1050080175 & 1634 $\pm$ 62 & 313.3 $\pm$ 2.11 & 6.8 & 26.0 \\
    27$^*$& 58350.685725 & 1050080177 & 1893 $\pm$ 66 & 326.3 $\pm$ 2.1 & 4.8 & 17.5 \\
    28$^*$& 58350.805694 & 1050080177 & 1765 $\pm$ 64 & 306.0 $\pm$ 2.0 & 4.5 & 18.5 \\
    $29^\diamond$& 58383.816006 & 1050080181 & 1556 $\pm$ 59 & 236.9 $\pm$ 4.5 & 6.0 & 31.0 \\
    30 & 58383.944918 & 1050080181 & 1606 $\pm$ 60 & 246.8 $\pm$ 2.3 & 7.5 & 18.5 \\
    31 & 58387.921859 & 1050080185 & 1607 $\pm$ 61 & 274.4 $\pm$ 2.4 & 5.0 & 24.5 \\
    32 & 58389.859359 & 1050080187 & 1487 $\pm$ 59 & 254.3 $\pm$ 2.3 & 9.0 & 23.5 \\
    33 & 58389.982440 & 1050080187 & 1503 $\pm$ 61 & 347.9 $\pm$ 6.2 & 4.25 & 21.5 \\
    34 & 58393.914340 & 1050080190 & 1505 $\pm$ 59 & 272.3 $\pm$ 2.3 & 5.3 & 30.0 \\
    35 & 58554.374787 & 2050080201 & 815 $\pm$ 46  & 248.2 $\pm$ 2.5 & 6.0 & 16.0 \\
    36 & 58554.891746 & 2050080201 & 1453 $\pm$ 58 & 260.1 $\pm$ 2.5 & 4.3 & 30.5 \\
    $37^\diamond$& 58563.592613 & 2050080207 & 1780 $\pm$ 60 & 75.3 $\pm$ 8.4 & 6.3 & 34.0 \\
    38 & 58596.953820 & 2050080209 & 1544 $\pm$ 61 & 318.9 $\pm$ 2.4 & 6.5 & 26.5 \\
    39 & 58597.386480 & 2050080210 & 1543 $\pm$ 61 & 315.9 $\pm$ 2.5 & 9.3 & 25.0 \\
    40 & 58597.391576 & 2050080210 & 499 $\pm$ 40  & 320.0 $\pm$ 2.6 & 5.5 & 11.5 \\
    41 & 58600.174653 & 2050080211 & 1565 $\pm$ 61 & 302.3 $\pm$ 2.5 & 5.5 & 26.5 \\
    42$^{+,\diamond}$ & 58600.732402 & 2050080211 & 1602 $\pm$ 61 & 261.3 $\pm$ 10.2 & 1.5 & 17.0 \\
    43$^+$ & 58600.740015 & 2050080211 & 331 $\pm$ 36  & 320.2 $\pm$ 2.5& 2.75 & 13.5 \\
    44$^{+,\diamond}$ & 58601.067781 & 2050080212 & 1520 $\pm$ 61 & 356.8 $\pm$ 2.4 & 9.0 & 16.5 \\
    45$^+$ & 58601.250281 & 2050080212 & 1596 $\pm$ 62 & 335.3 $\pm$ 2.5 & 8.0 & 19.0 \\
    46$^+$ & 58601.528921 & 2050080212 & 1277 $\pm$ 56 & 327.9 $\pm$ 2.3 & 3.0 & 12.0 \\
    47 & 58601.836030 & 2050080212 & 1516 $\pm$ 60 & 328.9 $\pm$ 2.5 & 8.8 & 20.0 \\
    $48^\diamond$& 58602.029559 & 2050080213 & 1561 $\pm$ 61 & 332.3 $\pm$ 2.2 & 5.0 & 26.0 \\
$49^\diamond$& 58602.034027 & 2050080213 & 1427 $\pm$ 59 & 348.2 $\pm$ 2.8 & 3.0 & 15.0 \\
    50$^+$ & 58602.151831 & 2050080213 & 1516 $\pm$ 60 & 318.9 $\pm$ 2.4 & 6.3 & 30.5 \\
    51 & 58602.350552 & 2050080213 & 1503 $\pm$ 60 & 322.4 $\pm$ 2.2 & 5.8 & 27.5 \\
    \enddata
    \tablenotetext{a}{ Peak count rates are pre-burst rate subtracted.}
        \tablenotetext{b}{ Data acquisition stopped before the e-folding time.}
    \tablenotetext{*} { Bursts were observed simultaneously by AstroSat.}
    \tablenotetext{+} { Bursts were observed simultaneously by NuSTAR.}
    \tablenotetext{\diamond} { These bursts are detected only within the unfiltered events.}
    \label{tab:burst_prop}
\end{deluxetable*}    

\subsection{Search for burst oscillations} 

We performed an extensive search for high frequency oscillations within the entire sample of 51 bursts. We used the Z$_{n=1}^2$ test \citep[that is the Rayleigh test,][]{1983A&A...128..245B} in a dynamic manner as follows: We constructed Z$_{1}^2$ powers using photons in the 1$-$10 keV energy band in a 2~s time window starting from 5~s prior to the onset of the burst until 50~s after with the steps of 0.5~s. The frequency interval of our interest ranged from 576~Hz to 586~Hz, centering at the known oscillation frequency of 581~Hz with a frequency step of 0.5 Hz. Furthermore, we repeated this search procedure using photons in the 3--10 keV band, using a search window size of 4~s, and a sliding step of 1~s. Overall, our searches were performed with all eight combinations of the energy interval, window length and sliding time step settings. 

We find no significant detection for burst oscillations in our search in the 1--10 keV band. On the other hand, in the 3--10 keV band, we identified a potential case (burst 13) in which our search within both 2~s and 4~s windows resulted in a brief episode of oscillations at 582 Hz with more than 99\% significance, assuming that Z$_{1}^2$ powers are distributed as $\chi^2$ with 2 degrees of freedom and number of trials as the number of independent powers in each search window. Note however that the episode of this oscillation starts about 30~s after the onset of the event, when the burst flux has already significantly declined. Therefore, we claim this as a tentative candidate for burst oscillations.

We also searched for burst oscillations in the entire AstroSat sample of 27 bursts in the entire LAXPC energy range using the same strategy described above. We find no evidence for high frequency oscillations in any of these bursts.

\subsection{Time Resolved Spectroscopy}

The soft X-ray sensitivity combined with the larger effective area of \nicer~ allows for an effective time resolved spectral study of thermonuclear X-ray bursts. Following previous studies \citep[see e.g.,][]{2022MNRAS.510.1577G} we extracted time resolved X-ray spectra for each burst with varying exposure times, which depends on the total observed count-rate from the source. To be able to catch the spectral evolution during the rise of the bursts we used 0.5~s as a fixed exposure time up to the peak moment and then allowed the exposure time to be longer. We also tested shorter exposure times like 0.25~s but that caused the statistical uncertainties of the spectral parameters to be too large. For each observation we estimated the background using the version 7b of \texttt{nibackgen3C50}
\footnote{\url{https://heasarc.gsfc.nasa.gov/docs/nicer/tools/nicer\_bkg\_est\_tools.html}} tool \citep{2021arXiv210509901R}. 
We removed the detector IDs 14 and 34 as these two FPMs are known to show episodes of high noise rates, rom our analysis and generated response and ancillary response files using the \texttt{nicerrmf} and \texttt{nicerarf} tools for each observation individually. 
Throughout the analysis we used \emph{ciao} \citep{2006SPIE.6270E..1VF}, \emph{sherpa} \citep{2001SPIE.4477...76F,2007ASPC..376..543D,2020zndo...3944985B} and custom written python scripts utilizing the \emph{Astropy} \citep{2013A&A...558A..33A,2018AJ....156..123A}, \emph{NumPy} \citep{van2011numpy}, \emph{Matplotlib} \citep{Hunter:2007}, and \emph{Pandas} \citep{mckinney-proc-scipy-2010} libraries to perform the spectral analysis. Only in the application of the reflection model we used XSPEC \citep{Arnaud1996}. We grouped each spectra to have at least 50 counts in each channel and performed the fits in the 0.5--10.0~keV range. The reported unabsorbed bolometric fluxes are calculated with \emph{sample\_flux} command within \emph{sherpa}, using 10$^4$ simulations around the best fit parameters within the 0.01--200~keV energy range. For the calculation of the unabsorbed fluxes during the application of the reflection model we used \emph{cflux} in XSPEC and report the 0.5--10.0~keV band. Uncertainties are reported at the 68\% confidence level unless noted otherwise.

We note that, because we search for bursts in the unfiltered data, during some of the cases the overshoot\footnote{https://heasarc.gsfc.nasa.gov/docs/nicer/analysis\_ threads/overshoot-intro/} rate showed significant variations just before, during or after a burst. Overshoot rates are recorded when an energetic charged particle passes through the silicon drift detector. These rates are a good indicator of the non-X-ray background of \nicer~and therefore they are recorded and used by the standard filtering. These variations affect mostly the high energy part of the spectra and did not allow us to constrain the pre-burst spectral model reliably, which is needed to be well determined and used as a background for the analysis of the bursts. We therefore did not perform time resolved spectral analysis for these bursts. In some cases it was possible to use post burst X-ray spectra as our background and we used these in our analysis. As a result it was only possible to perform time resolved spectral analysis of 40 X-ray bursts observed from \source. 
Before or after each burst we extracted X-ray spectra with an exposure time of 100~s and used these to characterize the persistent emission from the system. We first modeled these spectra with absorbed disk blackbody models plus a power-law or a blackbody components. However, based on the initial fit results we decided to use an absorbed disk blackbody plus a power-law model. Note that in some cases only a power-law model provided good enough fits and the addition of a disk blackbody component either did not improve the fit or the parameters of this component could not be constrained, and we therefore used only the power-law component for these bursts. These are indicated in \autoref{tab:pers}. We assume interstellar abundance along the line of sight to the source and use the \emph{tbabs} model \citep{2000ApJ...542..914W} to take into account the interstellar extinction in the soft X-rays. Initially we allowed the Hydrogen column density to vary in the different observations. We then calculated the error weighted average of the resulting values, which resulted in N$_{\rm {H}}$ = 4.4$\times$10$^{21}$ cm$^{-2}$, with a standard deviation of 0.5$\times$10$^{21}$ cm$^{-2}$. We used this average value as a fixed parameter for the whole analysis, which agrees with previously reported values \citep{2006AAS...209.1712J,2012ApJ...747...76G}. Results of the best fit models to these spectra are given in \autoref{tab:pers}.

For the time resolved spectral analysis of the bursts we followed three different approaches. Initially, we fit the spectra with only the persistent emission model where the parameters are fixed to the values given in \autoref{tab:pers}. When this resulted in bad statistics, due obviously to the burst emission itself, we added a blackbody component to take into account the burst emission. However, similar to previous examples from \nicer~ \cite[see e.g.,][]{2019ApJ...883...61J,2021ApJ...907...79B,2021MNRAS.506.2020D,2022MNRAS.510.1577G} such a model often was not sufficient to obtain a statistically acceptable fit especially at around the peak flux moments of the bursts. As a second alternative we then applied the so-called \fa method \citep{2013ApJ...772...94W, 2015ApJ...801...60W}, where the pre-burst emission is assumed not to change in shape but scaled-up to compensate for the observed excess emission beyond the thermal emission from the burst itself. In each case we allowed the \fa parameter to vary from unity we ran an f-test and only kept that parameter free if the chance probability of the change in statistic is less than 5\%. The effect of the \fa parameter in two X-ray spectra~(Bursts 6 and 23) are shown in \autoref{fig:spec_var}. The best fit values for the scaling factor are \fa = 6.54$\pm$0.4 for burst 6 and \fa = 1.8$\pm$0.3 for burst 23. We also show the persistent emission model without the application of the \fa parameter in the same figure for comparison.  
\autoref{fig:bursts_plts_faint} shows time evolution of spectral parameters for some example X-ray bursts. 

\begin{deluxetable*}{ccccccc}
\tablehead{ \colhead{BID} & \colhead{T$_{in}$} & \colhead{Norm$_{\rm{DBB}}$ } & \colhead{$\Gamma$}  &\colhead{Norm$_{\rm{POW}}$ } & \colhead{Flux\tablenotemark{a}} & \colhead{$\chi^2$ / dof} \\
\colhead{} &  \colhead{(keV)} & \colhead{(R$^2_{\rm km}/\rm{D}_{10 {\rm kpc}}^{2}$)} & & & }

\tablecaption{Best fit model results for pre-burst X-ray spectra of \source using single power-law model or an absorbed disk blackbody plus a power-law model.\label{tab:pers}}

\startdata
2&0.98$\pm$0.01&87$\pm$9&1.47$\pm$0.03& 0.34$\pm$0.01& 4.32$\pm$0.28&417.2/454\\
3&0.72$\pm$0.02&148$\pm$18 &1.63$\pm$0.03& 0.24$\pm$0.01&2.33$\pm$0.14&371.7/352\\	
5&0.90$\pm$0.03& 71$\pm$10 &1.41$\pm$0.03&0.24$\pm$0.01&2.98$\pm$0.21&395.9/401\\	
6&0.72$\pm$0.02& 159$\pm$19 &1.64$\pm$0.03&0.23$\pm$0.01&2.34$\pm$0.18&358.8/348\\
7&0.71$\pm$0.02&163$\pm$19 &1.60$\pm$0.01&0.23$\pm$0.01& 2.40$\pm$0.16&353.5/356\\	
9&0.39$\pm$0.05&149$\pm$114 &1.79$\pm$0.04&0.18$\pm$0.01&1.12$\pm$0.09&258.0/258\\
11&1.84$\pm$0.12&1.5$\pm$0.4&2.06$\pm$0.03&0.16$\pm$0.01&1.06$\pm$0.14&259.9/258\\
12&--\tablenotemark{b}&--&1.89$\pm$0.01&0.19$\pm$0.01&1.01$\pm$0.02&323.0/255\\	
13&--&--&1.90$\pm$0.01&0.18$\pm$0.02&0.96$\pm$0.01&260.6/246\\
14&0.64$\pm$0.03&109$\pm$21&1.68$\pm$0.06&0.14$\pm$0.01&1.24$\pm$0.11&266.9/266\\
15&0.81$\pm$0.01&180$\pm$13&1.22$\pm$0.03& 0.16$\pm$0.01&3.22$\pm$0.26&366.7/400\\
16&0.85$\pm$0.02&110$\pm$12&1.54$\pm$0.03&0.24$\pm$0.01&2.81$\pm$0.21&357.4/385\\
18&0.65$\pm$0.02&174$\pm$25&1.71$\pm$0.02&0.22$\pm$0.01&1.95$\pm$0.14&353.8/321\\
19&--& --&1.72$\pm$0.01& 0.29$\pm$0.02&1.77$\pm$0.02&366.6/318\\
20&0.54$\pm$0.02 & 252$\pm$50 &1.70$\pm$0.02&0.28$\pm$0.01 &2.09$\pm$0.13&298.0/328\\
21&--&--&1.81$\pm$0.01& 0.20$\pm$0.01 &1.12$\pm$0.01&335.0/270\\	
22&2.72$\pm$0.02&0.6$\pm$0.2 &2.10$\pm$0.03&0.20$\pm$0.01 &1.42$\pm$0.24&294.1/287\\
23&--& --&1.77$\pm$0.01& 0.24$\pm$0.01&1.41$\pm$0.02&373.4/288\\
24&2.67$\pm$0.25& 0.42$\pm$0.22 &1.96$\pm$0.08& 0.24$\pm$0.01 &1.55$\pm$0.25&337.2/298\\
25&--& --&1.76$\pm$0.02& 0.20$\pm$0.02 &1.17$\pm$0.02&195.0/210\\
26&--& --&1.79$\pm$0.01& 0.24$\pm$0.01 &1.42$\pm$0.01&324.3/290\\	
27&--& --&1.77$\pm$0.01& 0.25$\pm$0.01 &1.50$\pm$0.02&329.2/298\\
28&--& --&1.74$\pm$0.01& 0.23$\pm$0.01 &1.41$\pm$0.02&331.8/290\\	
29&3.26$\pm$0.25&0.5$\pm$0.2 &1.86$\pm$0.02& 0.09$\pm$0.01 &1.28$\pm$0.39&303.5/270\\
30&3.58$\pm$0.30&0.3$\pm$0.1 &1.77$\pm$0.02& 0.11$\pm$0.01&1.36$\pm$0.36&277.8/291\\
32&2.81$\pm$0.12&0.8$\pm$0.2  &1.95$\pm$0.18&0.09$\pm$0.01&1.35$\pm$0.30&292.3/291\\
34&2.92$\pm$0.16&0.8$\pm$0.2 &2.10$\pm$0.18&0.10$\pm$0.01 &1.44$\pm$0.30&335.1/298\\
35&3.01$\pm$0.25&0.5$\pm$0.2  &1.65$\pm$0.16& 0.11$\pm$0.01&1.36$\pm$0.29&278.4/288\\
36&3.09$\pm$0.20&0.6$\pm$0.2 &1.79$\pm$0.14&0.11$\pm$0.01&1.42$\pm$0.29&258.1/296\\
38&--&--&1.57$\pm$0.01&0.22$\pm$0.01&1.63$\pm$0.02&324.8/314\\
39&--&--&1.53$\pm$0.01&0.22$\pm$0.01&1.65$\pm$0.02&394.9/312\\
40&--&--&1.53$\pm$0.01&0.22$\pm$0.01&1.67$\pm$0.02&397.3/318\\
41&--&--&1.52$\pm$0.01&0.21$\pm$0.01&1.58$\pm$0.02&377.1/305\\
42&0.19$\pm$0.02 & 7055$\pm$4320&1.41$\pm$0.02& 0.20$\pm$0.01&1.84$\pm$0.40&358.4/324\\	43&0.19$\pm$0.02&7974$\pm$463&1.45$\pm$0.02&0.20$\pm$0.01&1.73$\pm$0.08&333.5/313\\
45&--&--&1.59$\pm$0.01&0.24$\pm$0.01&1.68$\pm$0.02&415.4/322\\
46&--&--&1.64$\pm$0.01&0.24$\pm$0.01&1.62$\pm$0.02&330.5/311\\
47&--&--&1.56$\pm$0.01&0.23$\pm$0.01&1.68$\pm$0.02&383.3/319\\
50&--&--&1.58$\pm$0.01&0.22$\pm$0.01&1.61$\pm$0.02&336.2/310\\
51&--&--&1.56$\pm$0.01&0.22$\pm$0.01&1.64$\pm$0.02&400.4/311\\
\enddata												
\tablenotetext{a}{Unabsorbed 0.5$-$10~keV flux in units of $\times10^{-9}$~\fluxcgs.}
\tablenotetext{b}{For these bursts a disk blackbody component is not required.}
\end{deluxetable*}    

To account for the soft excess emission during the bursts, instead of the \fa method, we applied a model where the reflection of the burst emission off of the accretion disk is taken into account. For this purpose, we used a specific version of the \texttt{bbrefl}\footnote{https://heasarc.gsfc.nasa.gov/xanadu/xspec/models/bbrefl.html} model \citep{2004ApJ...602L.105B, 2004MNRAS.351...57B} in XSPEC \citep{Arnaud1996}, which takes into account the reflection of thermal photons from an accretion disk. We also took into account the general relativistic effects in the inner disk using a convolution model~\texttt{relconv} \citep{2010MNRAS.409.1534D,2022ApJ...926...13G}. The \texttt{bbrefl} model depends on the ionization parameter ($\log \xi$) in the disk, the temperature of the incident blackbody and iron abundance component in the accretion disk. Here the ionization parameter is defined as the ratio of the incident flux to the hydrogen number density in the disk. The \texttt{relconv}\footnote{https://heasarc.gsfc.nasa.gov/xanadu/xspec/manual/node297.html} model depends on a large number of parameters, like the inner and outer emissivity indices, the dimensionless spin parameter~(a), inner and outer radii of the disk as well as the orbital inclination.  We applied a few reasonable conditions when using the \texttt{bbrefl} and \texttt{relconv}. First, we tied the inner emissivity index to outer emissivity index. Next, we fix the dimensionless spin parameter~(a), in the \texttt{relconv} model to 0.27 in the fits, assuming typical values for a neutron star  (with a mass of 1.4M$_{\sun}$ and a radius of 10~km) of the neutron star and a spin period of 581~Hz which contains the approximation for converting spin frequency to dimensionless spin \citep{2000ApJ...531..447B}. Also, we fixed the inner and outer radius of the disk at R$_{ISCO}$ and 1000.0 R$_{g}$, respectively.  We also assumed that the orbital inclination of the system is $60^{\circ}$ degrees \citep{2006MNRAS.373.1235C}. Moreover, when using the reflection model, the ionization parameter ($\xi$) is fixed at 3.4. As a test, we allow the ionization parameter to vary, however, the fit does not improve in the majority of cases and the parameter tends to maximally ionized. While fitting the spectra with this model we linked the incident temperature in the reflection model to the burst temperature. We fixed the hydrogen column density at 4.4$\times$10$^{21}$ cm$^{-2}$ for all fits. We note that we only used the reflection model where the use of an \fa parameter greater than 1 is the statistically preferred model and the burst flux is larger than half the peak flux of the burst. 

Finally, taking advantage of hard X-ray sensitivity of AstroSat/LAXPC, we extracted X-ray spectra for the exact time intervals of the \nicer~ spectra for the five simultaneous bursts. This way it was possible to perform simultaneous soft and hard X-ray spectral fitting for these bursts. We note that, adding the 3.0--25~keV data to the 0.5--10~keV data from \nicer~ allows for a better determination of the high energy tail of the pre-burst emission so the inferred best fit spectral parameters of the pre-burst emission show some variations. For this reason, in \autoref{tab:astro_pers} we provide the best fit parameters for the NICER+LAXPC data we used to fit to the burst spectra. For the rest of the time resolved spectroscopy of \nicer~ and LAXPC data we followed the exact same methods as we did for only \nicer~ data.
 
 \begin{deluxetable*}{ccccccc}
\tablehead{ \colhead{BID} & \colhead{T$_{in}$} & \colhead{Norm$_{\rm{DBB}}$ } & \colhead{$\Gamma$}  & \colhead{Flux\tablenotemark{a}} & \colhead{Flux\tablenotemark{b}} & \colhead{$\chi^2$ / dof} \\
\colhead{} &  \colhead{(keV)} & \colhead{(R$^2_{\rm km}/\rm{D}_{10 {\rm kpc}}^{2}$)} & & }
\tablecaption{Best fit model results for pre-burst X-ray spectra of \source using an absorbed disk blackbody plus a power-law model, when \nicer~ and AstroSat/LAXPC is taken into account. Throughout the analysis we fixed neutral hydrogen column density to N$_{\rm{H}}$= 4.4$\times10^{21}~\rm{cm}^{-2}$. \label{tab:astro_pers}}

\startdata
7\tablenotemark{c}&0.92$\pm$0.01&83$\pm$5.0&2.01$\pm$0.02&1.17$\pm$0.01&1.18$\pm$0.01&353.52/356\\
22&1.87$\pm$0.11&1.0$\pm$0.2&1.90$\pm$0.01&0.76$\pm$0.01&1.02$\pm$0.02&294.10/287\\
23&	2.58$\pm$0.20&0.26$\pm$0.1&1.91$\pm$0.01 &0.79$\pm$0.01&1.04$\pm$0.01&319.70/286\\
27&	2.13$\pm$0.13&0.6$\pm$0.1&1.92$\pm$0.01&0.83$\pm$0.01&1.06$\pm$0.01&280.59/296\\
28&	2.35$\pm$0.17&0.4$\pm$0.1&1.92$\pm$0.01&0.80$\pm$0.01&1.10$\pm$0.02&305.76/288\\
\enddata									
\tablenotetext{a}{Unabsorbed 0.5$-$10.0~keV flux in units of $\times10^{-9}$~\fluxcgs.}
\tablenotetext{b}{Unabsorbed 3.0$-$25.0~keV flux in units of $\times10^{-9}$~\fluxcgs.}
\tablenotetext{c}{For this particular burst a Gaussian line at 6.5~keV was also added.}
\end{deluxetable*}    

\section{Results}
\label{sec:results}

\subsection{General Characteristics of Detected X-ray Bursts}
\label{sec:general_char}

\begin{figure*}
    \centering
    \includegraphics[scale=0.55]{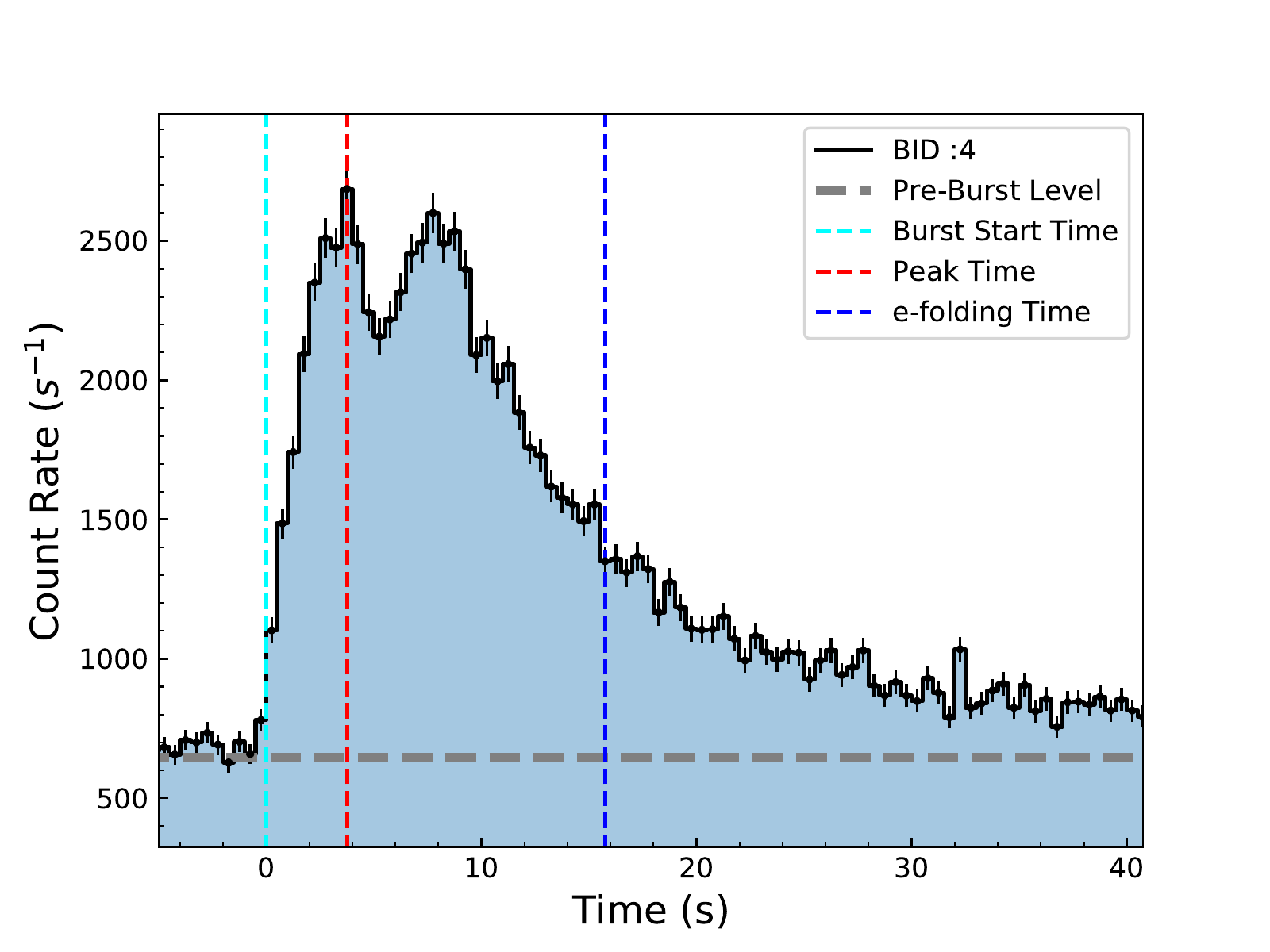}
    \includegraphics[scale=0.55]{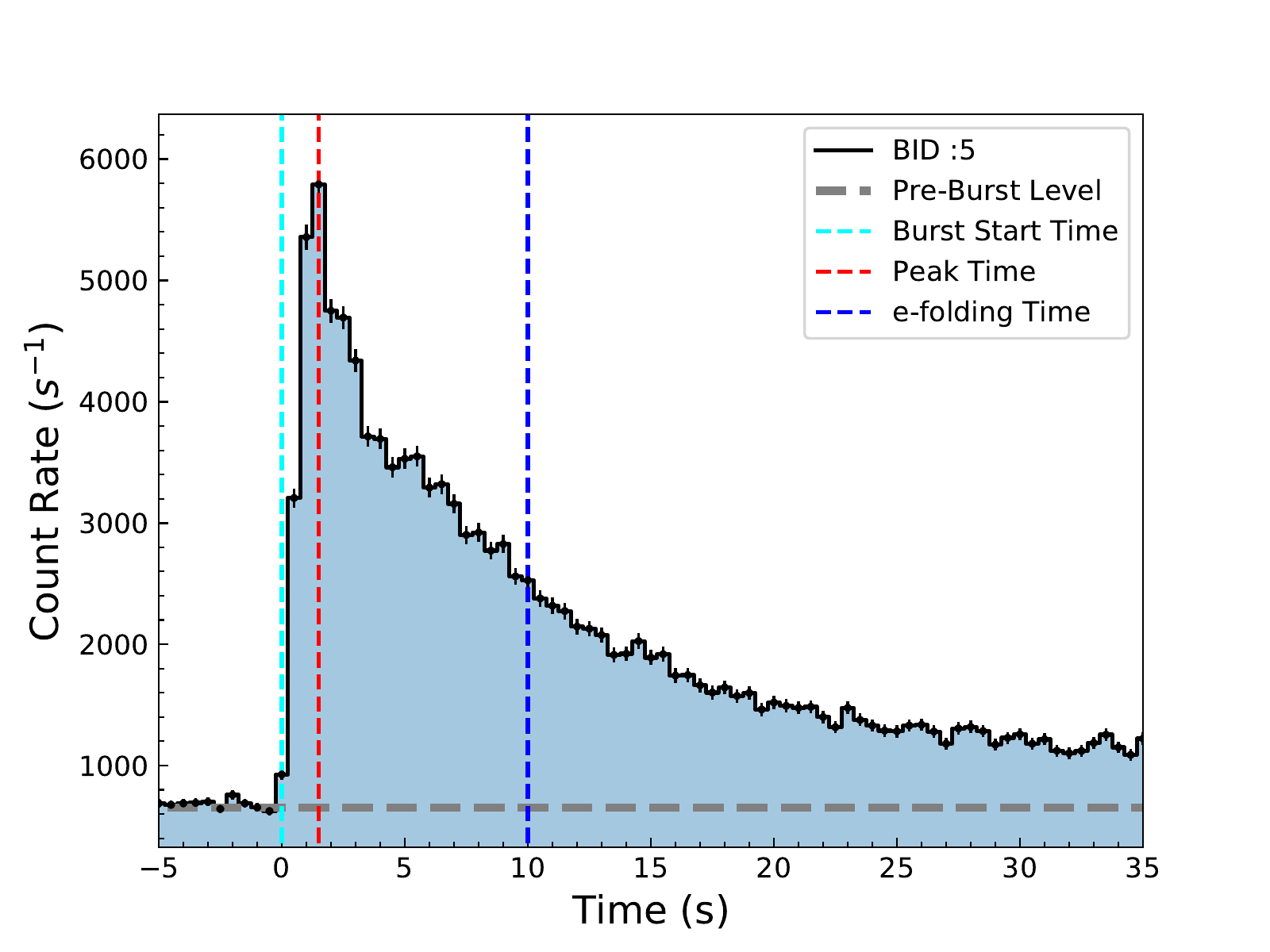}
    \includegraphics[scale=0.55]{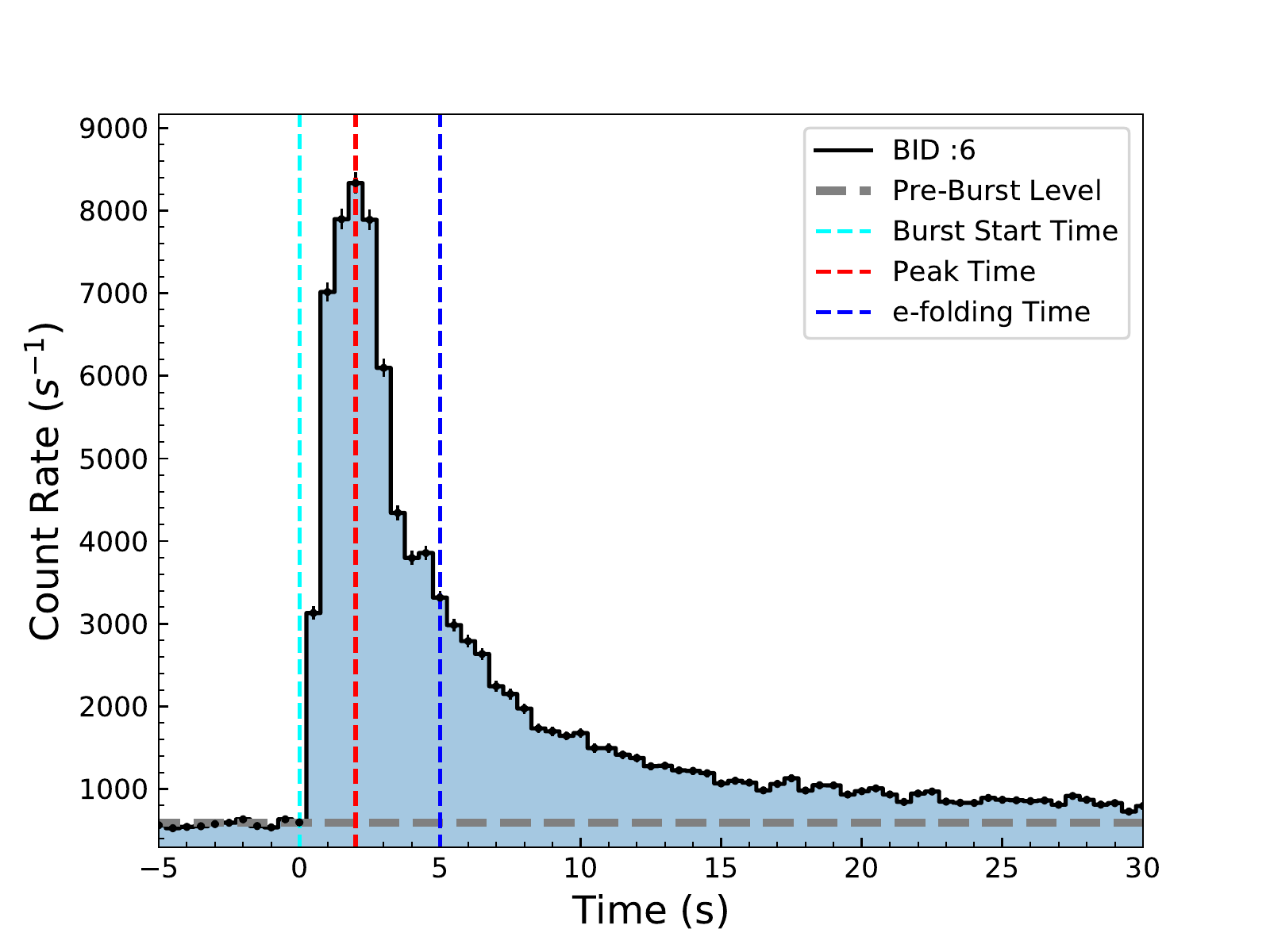}
    \includegraphics[scale=0.55]{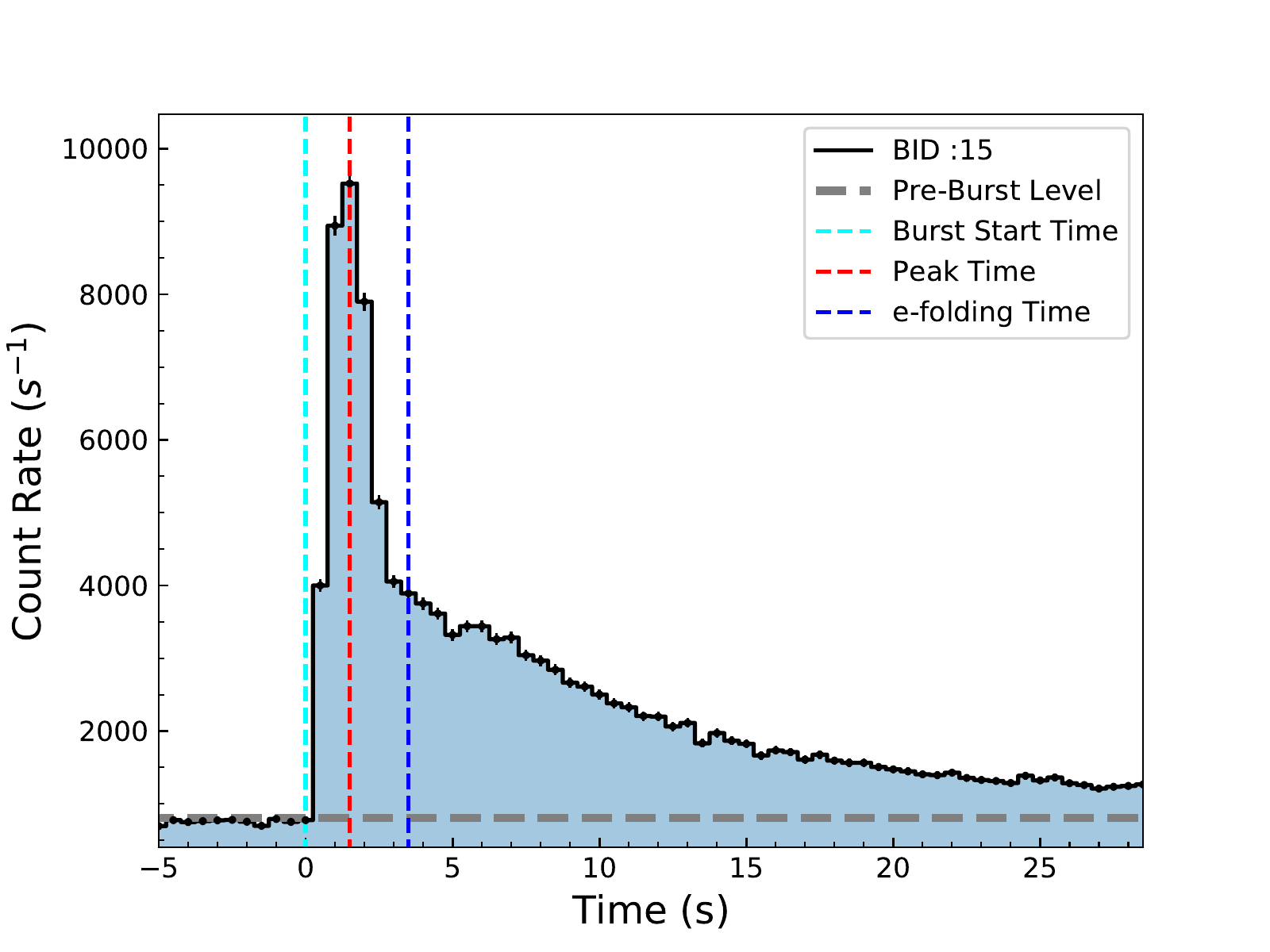}
    \includegraphics[scale=0.55]{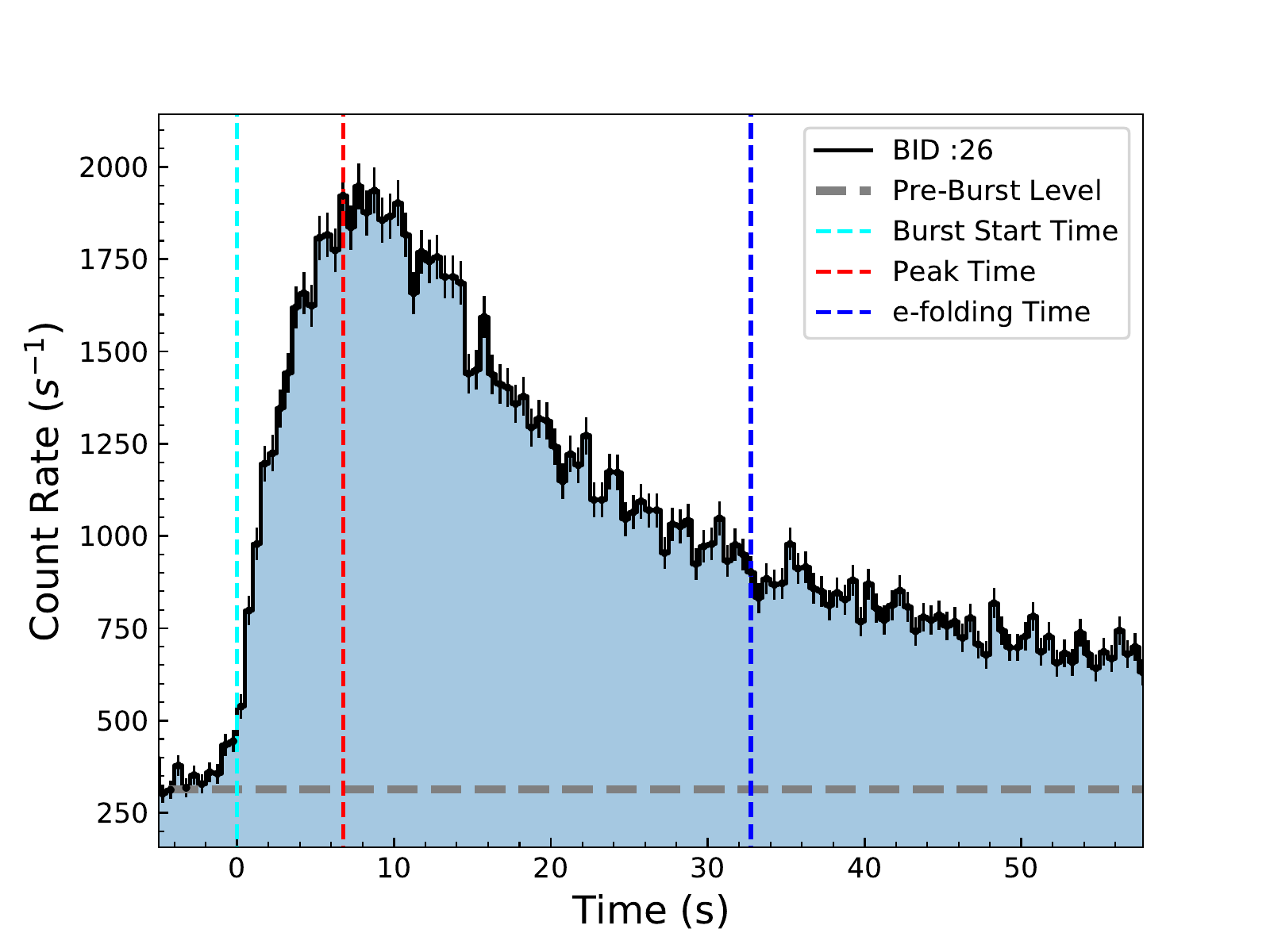}
    \caption{0.5-10~keV lightcurves of some of the X-ray bursts observed from \source. Bursts shown here are bursts 4, 5, 6, 15, and 26. Bursts 5, 6, and 15 are among the brightest bursts within the \nicer~ sample, while burst 4 is unique in having a double peaked structure and finally burst 26 is an example of a more typical X-ray burst.}
    \label{fig:example_lc}
\end{figure*}

Within the 520~ks long observations we detected roughly one burst per 10~ks resulting in a burst rate of 0.35~hr$^{-1}$, which is a little higher but broadly in agreement with the burst rate derived from previous missions for \source \citep{Galloway2008, 2020ApJS..249...32G}. Lightcurves of a few example bursts are shown in  \autoref{fig:example_lc}. As evident in \autoref{fig:example_lc}, burst 4 shows a pronounced double peaked structure in its lightcurve. Unfortunately during this burst the overshoot rate, which is generally caused by charged particle events, showed a continuous increase, which although does not change the reality of the bursts, prevented a detailed spectral analysis. 

\label{app:example_lc}

Some of the basic properties of the bursts observed with \nicer~ are given in \autoref{tab:burst_prop}. Within the sample three bursts quickly stand out from the rest with significantly higher peak count rates. These are the bursts 5, 6, and 15 with 5140$\pm$107, 7741$\pm$129, and 8715$\pm$138 count/s, respectively and their lightcurves are shown in \autoref{fig:example_lc}. These bursts also show evidence for photospheric radius expansion (see \autoref{sec:time_res}). The somewhat low peak flux of burst 5 compared to 6 and 15 suggests that it was bursting in a hydrogen-rich rather than a hydrogen-poor environment (see also \autoref{tab:time_res_results}). Besides these brightest bursts we note that there are seven bursts (bursts 7, 10, 13, 25, 35, 40, and 43) with peak count rates less than 1000 count/s. Four of these bursts are the secondary or tertiary bursts in short recurrence burst events. These are bursts 10, 13, 40, and 43. For bursts 25 and 35, our observations do not show evidence for a prior burst within ~69 and 170 s respectively, which is not very constraining, so these bursts may still be secondary bursts of short recurrence events that \nicer~ simply missed, given the typical recurrence time in such events as 720~s \citep{2010ApJ...718..292K}. However, for burst 7 we note that there is no prior burst within 1640~s. Implying that this burst may actually be intrinsically dimmer than a typical burst. The remaining bursts have an average of 1700 count/s peak rate with a standard deviation of 250 count/s and have pretty much similar profiles to each other. An example is shown in \autoref{fig:example_lc}.

Within the sample of 51 bursts there are a number of short recurrence events. We discuss these events and show the lightcurves in the \autoref{appendix_section}. 

In total we identified 15 X-ray bursts in the \nustar~ observation that spans 195~ks with 92~ks on-source exposure time. A detailed analysis of all of the bursts is beyond the scope of this study however, we would note that within this sample, we identified two short recurrence events with recurrence times of only 230~s and 629~s and one triple burst event where the last two of these bursts are also observed simultaneously with \nicer~ (see \autoref{fig:nustar_burst_lc}). Within the total \nustar~ burst sample,  7 bursts (burst 1, 5, 9, 10, 11, 14, and 15 within this observation) showed very similar profiles in their lightcurve with peak a count rate of 300 counts/s. We used these bursts to search for evidence for Compton cooling.

\subsection{Time Resolved Spectroscopy}
\label{sec:time_res}
\begin{figure*}
    \centering
    \includegraphics[scale=0.5]{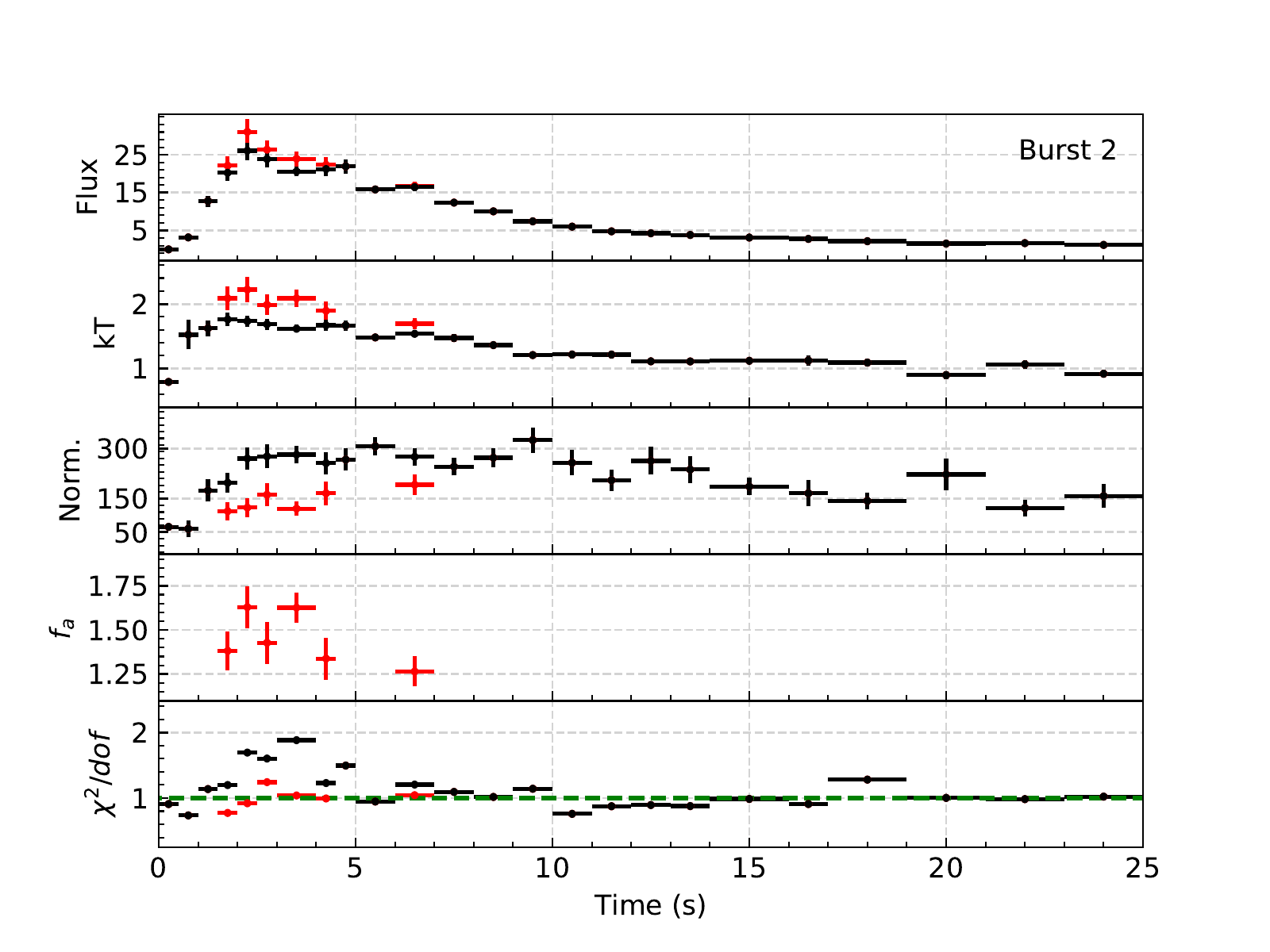}
    \includegraphics[scale=0.5]{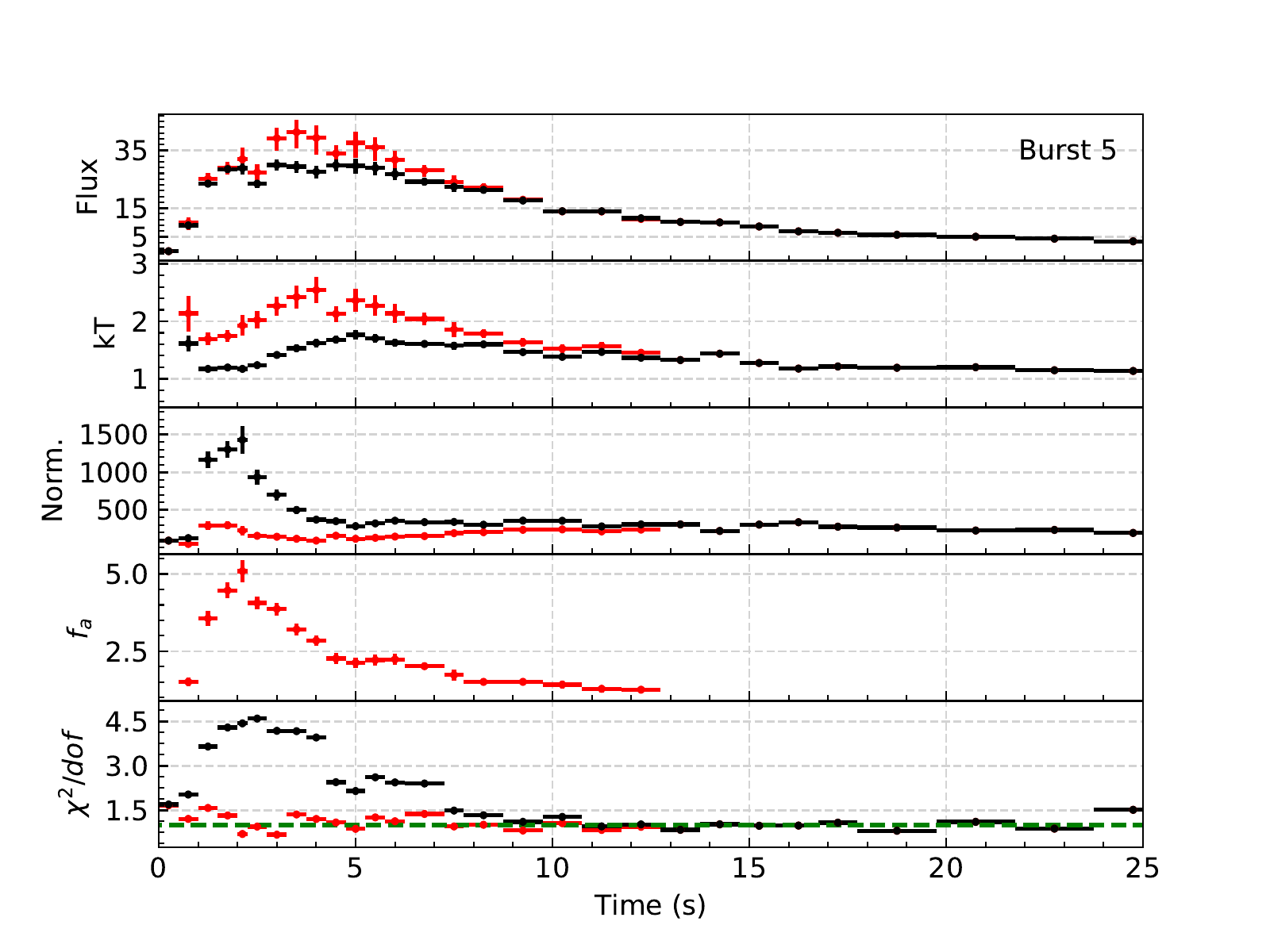}
    \includegraphics[scale=0.5]{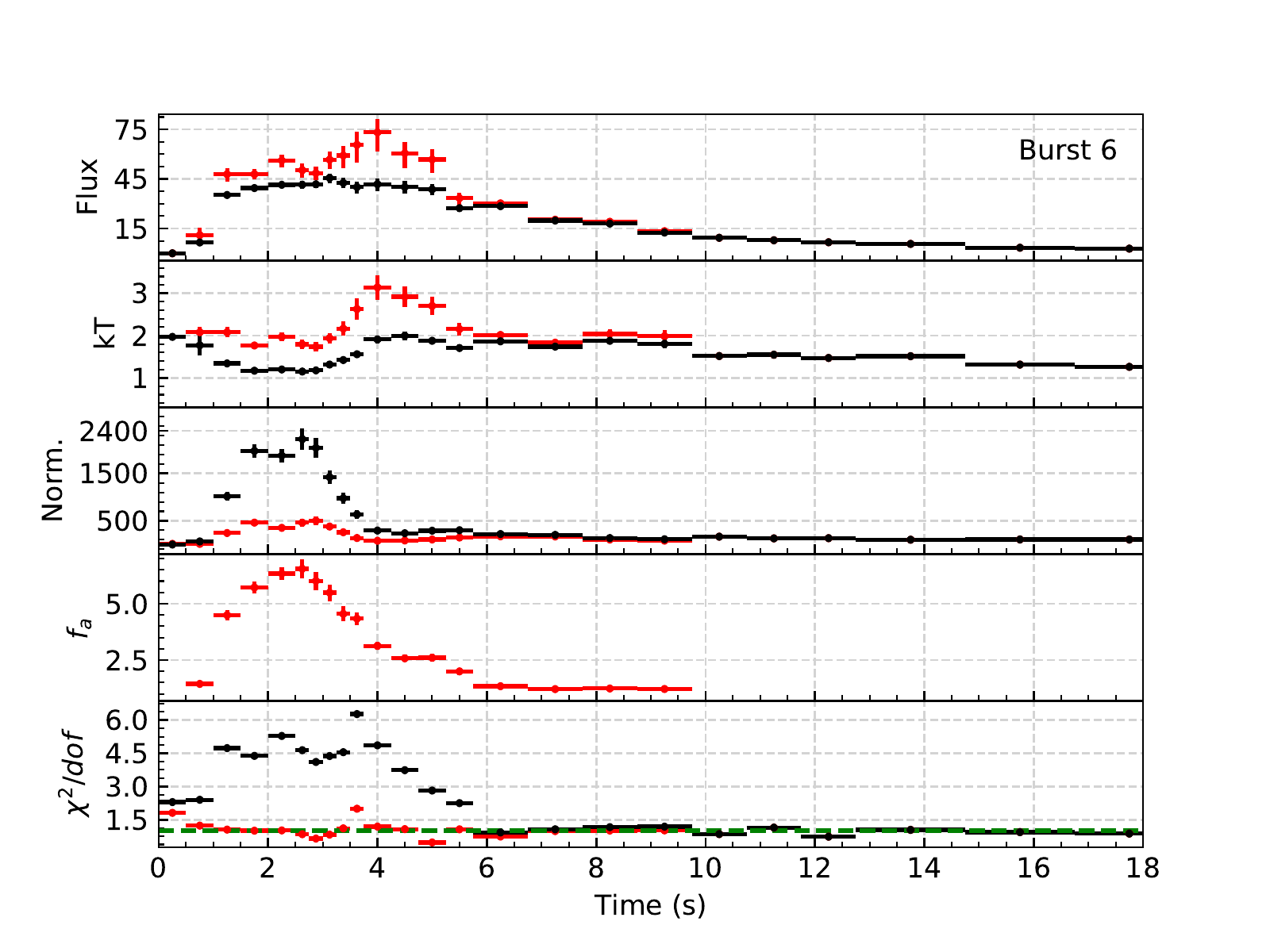}
    \includegraphics[scale=0.5]{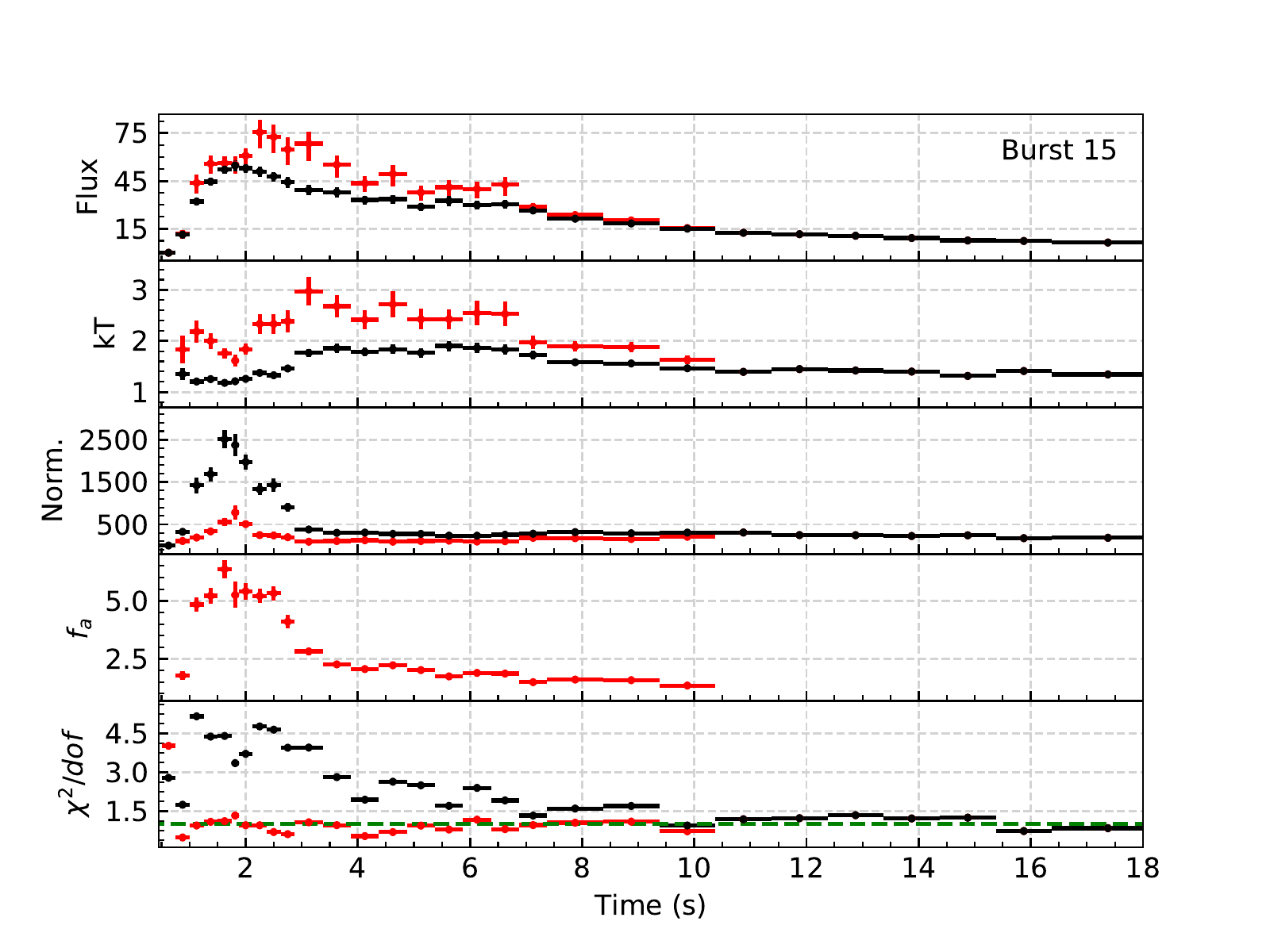}
    \caption{The time evolution of the NICER spectral parameters is shown for bursts 2, 5, 6 and 15. Red symbols show the results of $f_a$ method and black symbols show the results for constant background emission. From top to bottom, we show bolometric flux (in units of 10$^{-9}$~\fluxcgs), temperature (keV), blackbody normalization ($\rm{R}^{2}_{km}/\rm{D}^{2}_{10kpc}$), $f_a$, and finally the fit statistic, respectively.}
    \label{fig:bursts_plts_faint}
\end{figure*}

\begin{figure*}
    \centering
    \includegraphics[scale=0.5]{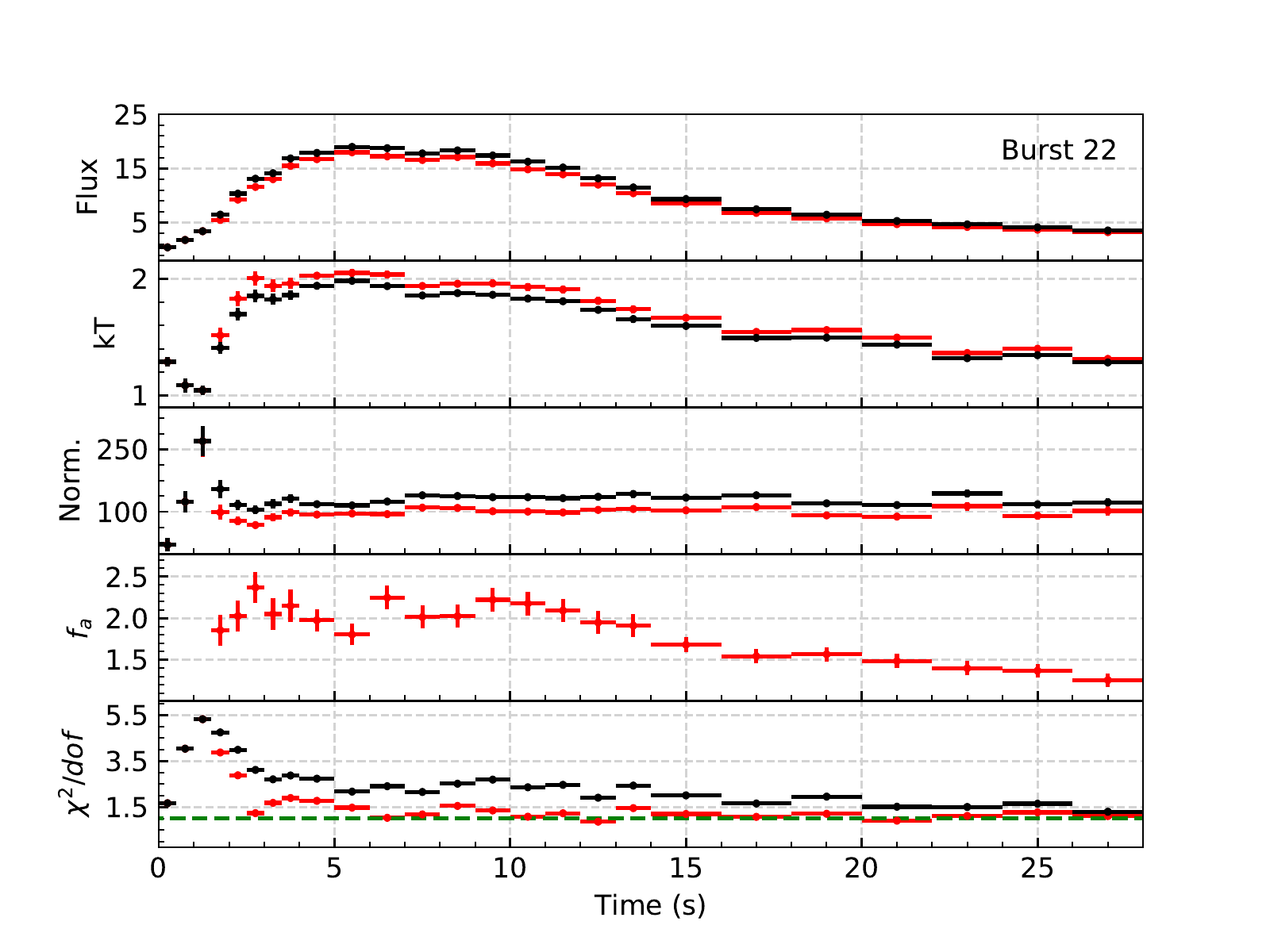}
    \includegraphics[scale=0.5]{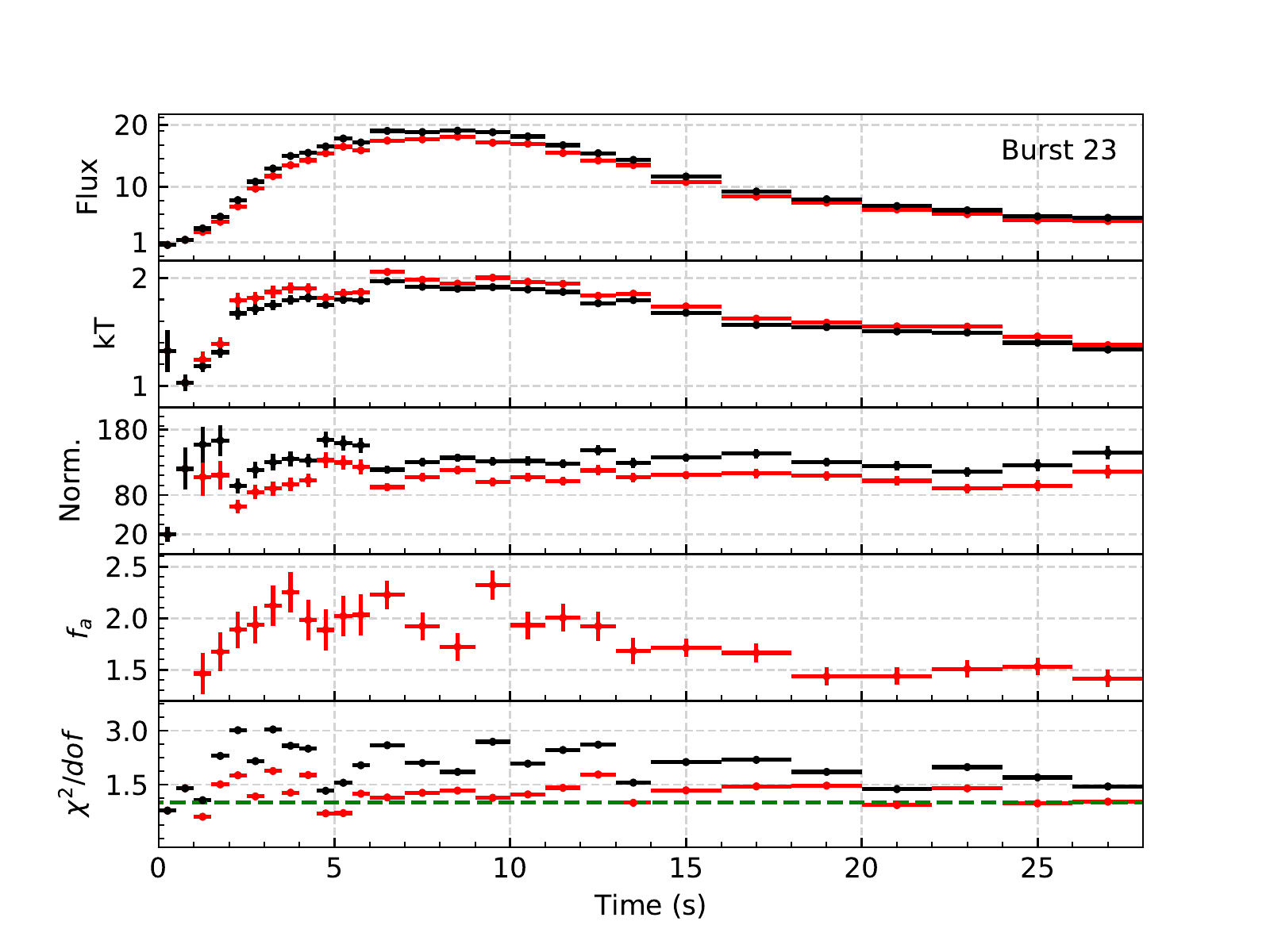}
    \caption{The time evolution of the NICER and AstroSat/LAXPC spectral parameters is shown for bursts 22 and 23. Red symbols show the results of $f_a$ method and black symbols show the results for constant background emission. We show in order from top to bottom, 0.5--10~keV X-ray flux (in units of 10$^{-9}$~\fluxcgs), temperature (keV), blackbody normalization ($\rm{R}^{2}_{km}/\rm{D}^{2}_{10kpc}$), $f_a$, and finally the fit statistic, respectively. }
    \label{fig:astrosat_bursts_plts}
\end{figure*}

\begin{figure}
    \centering
    \includegraphics[scale=0.5]{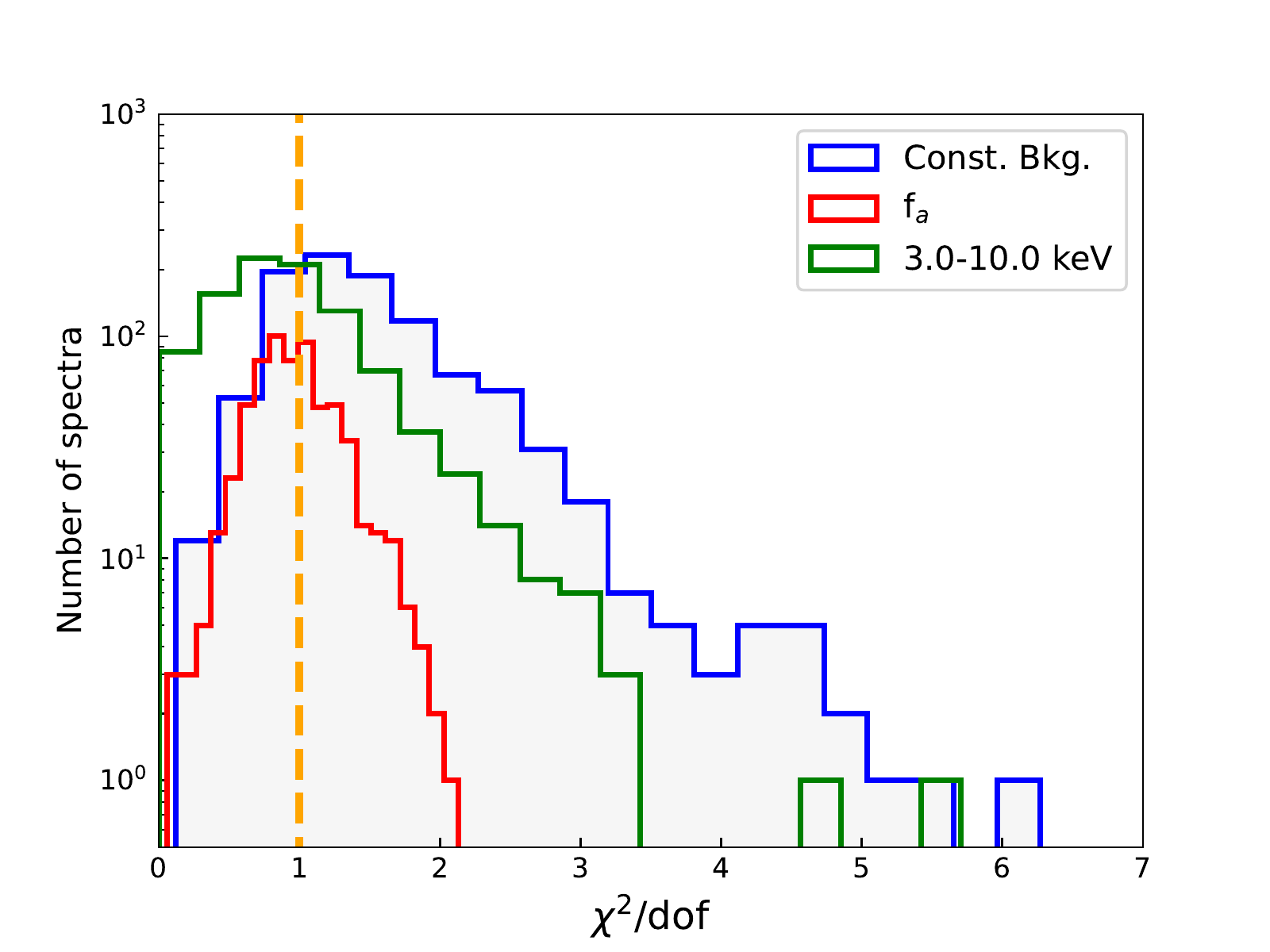}
    \caption{Histogram of $\chi^2$/dof values for the 0.5--10.0 keV (blue histogram), 3.0--10~keV (green histogram) constant background approach fits, as well as the 0.5--10~keV fits where the scaling factor \fa is used for the pre-burst emission (red histogram).}
    \label{fig:chi_hist}
\end{figure}

\begin{figure}
    \centering
    \includegraphics[scale=0.3]{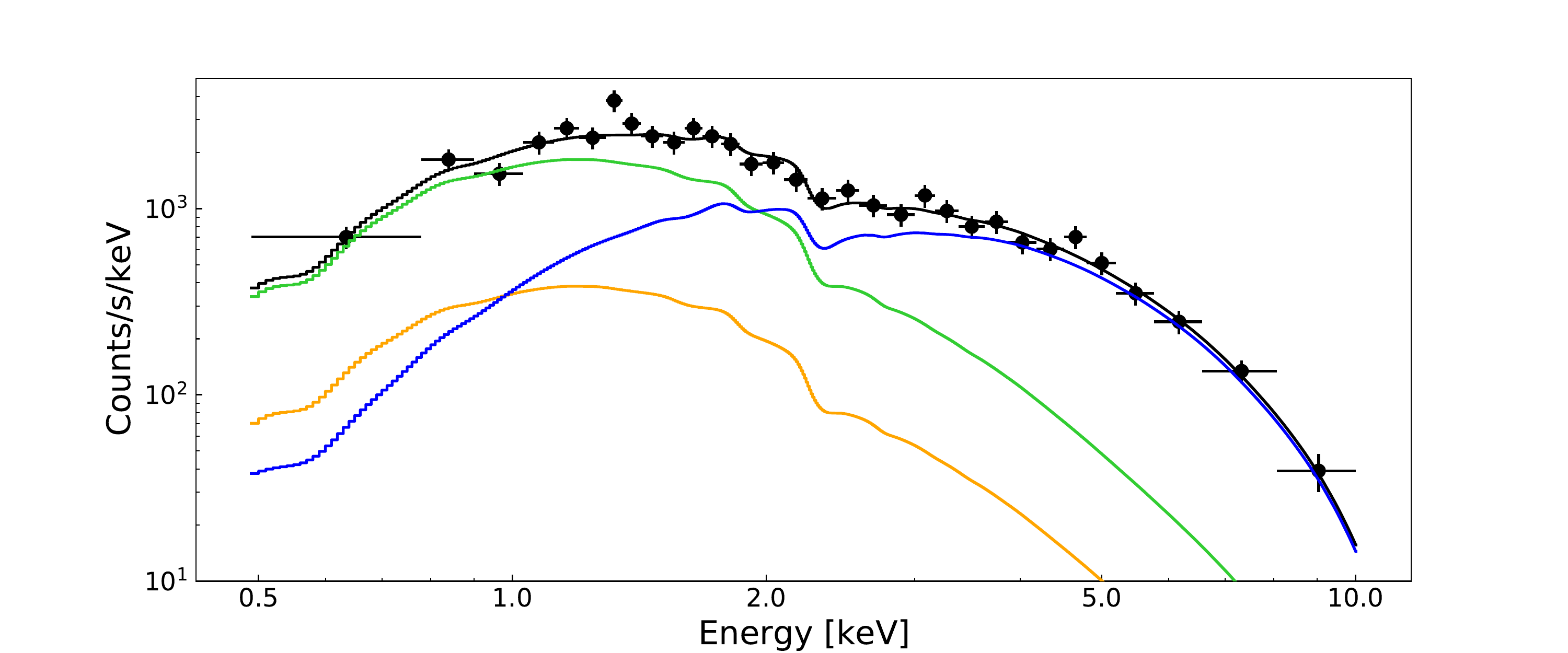}
        \includegraphics[scale=0.3]{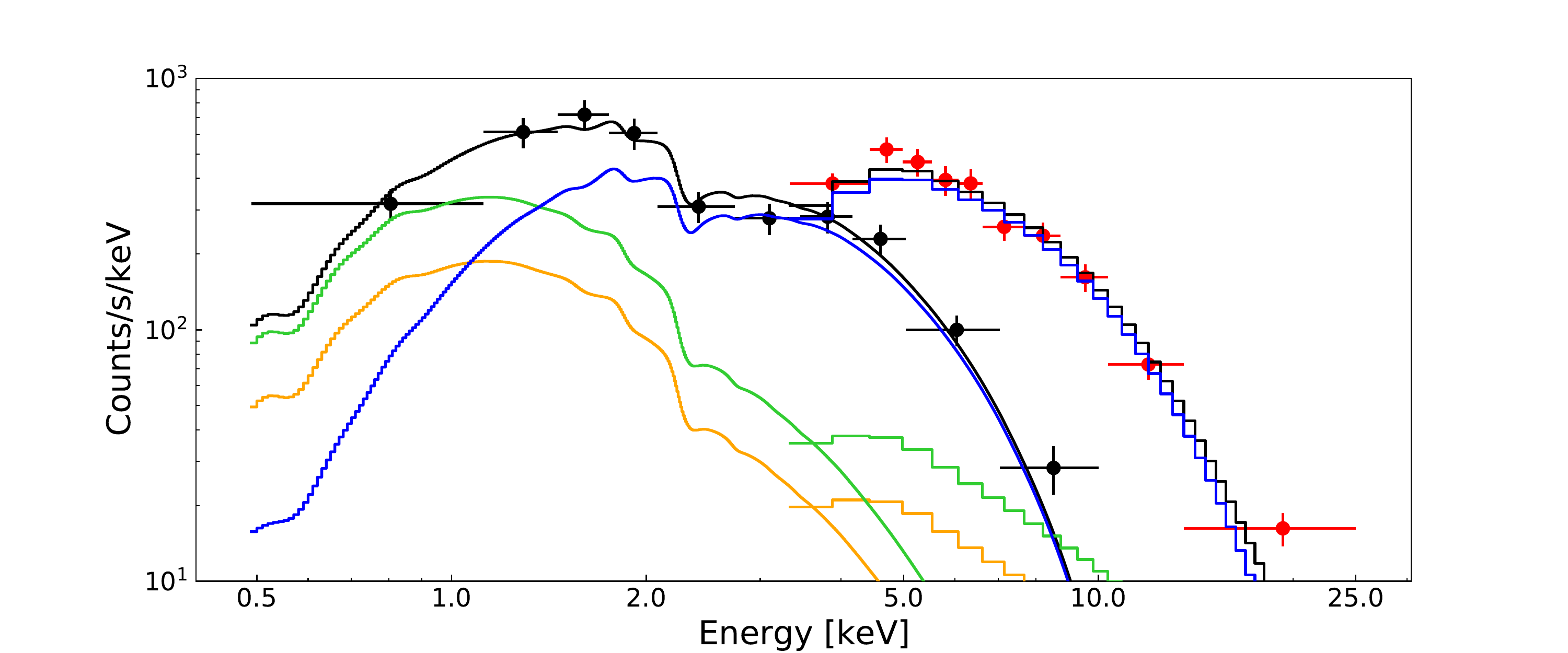}
    \caption{Examples of best fit X-ray spectra extracted around the peak flux moment of Burst 6 and burst 23 is shown in the upper and lower panel, respectively. In both panels black and red dots show the observed actual NICER and LAXPC data, respectively. Black lines show the summed best fit models. Blue and green lines show the individual components of the models, which are the blackbody model for the burst and the persistent emission model multiplied with the \fa parameter, respectively. Also, in orange without the application of the \fa parameter in both panels.
    }
    \label{fig:spec_var}
\end{figure}

\begin{figure}
    \centering
    \includegraphics[scale=0.5]{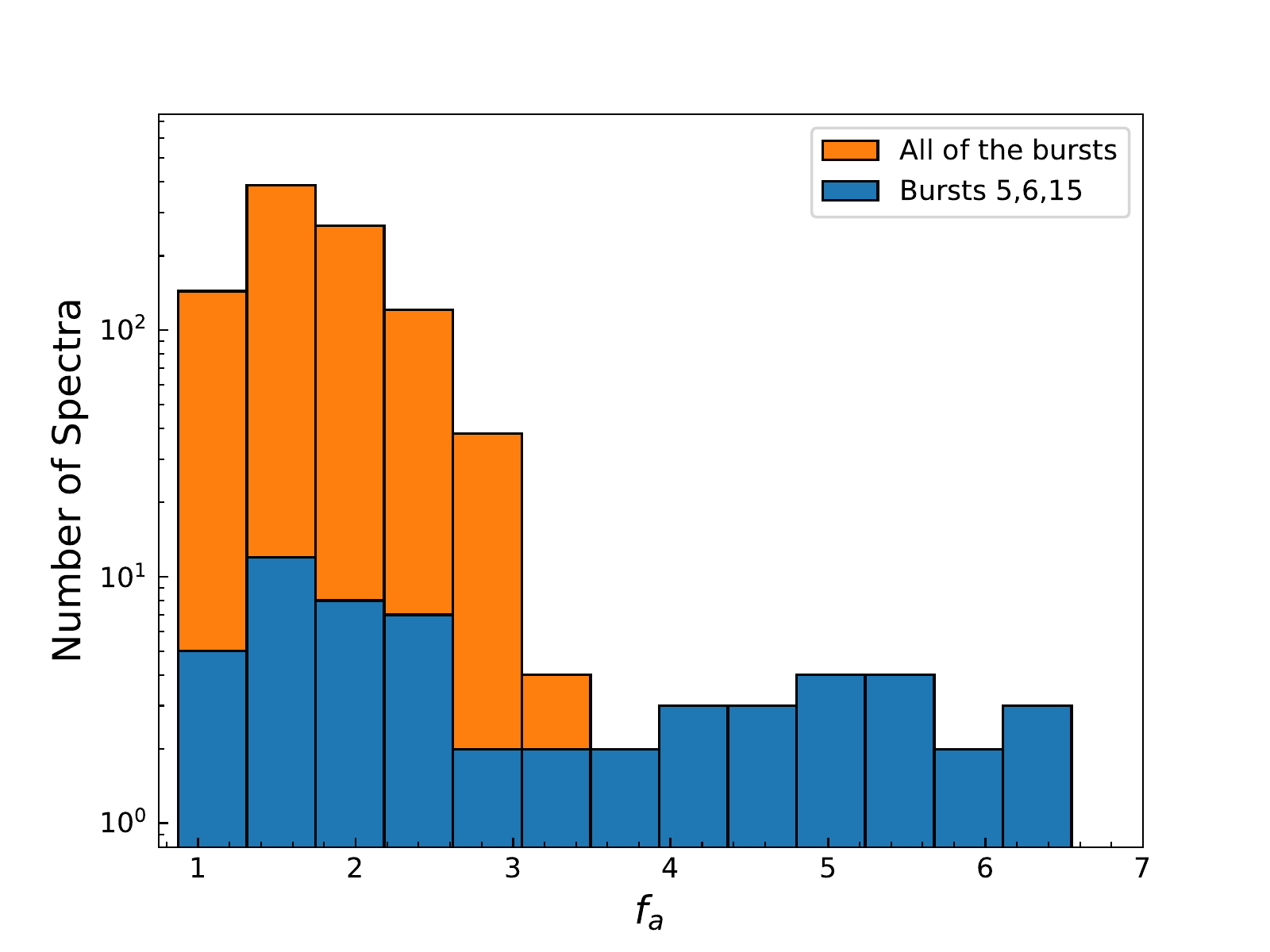}
    \caption{Histogram of the best fit \fa values, where the burst flux is greater than 10$^{-9}$\fluxcgs is shown in orange. With blue color the histogram of \fa values for only bursts 5, 6, and 15 are shown which are suspected to show photospheric radius expansion. Note that due to difference in sizes of the samples the number of bins are kept different.}
    \label{fig:fa_hist}
\end{figure}

\begin{figure}
    \centering
    \includegraphics[scale=0.5]{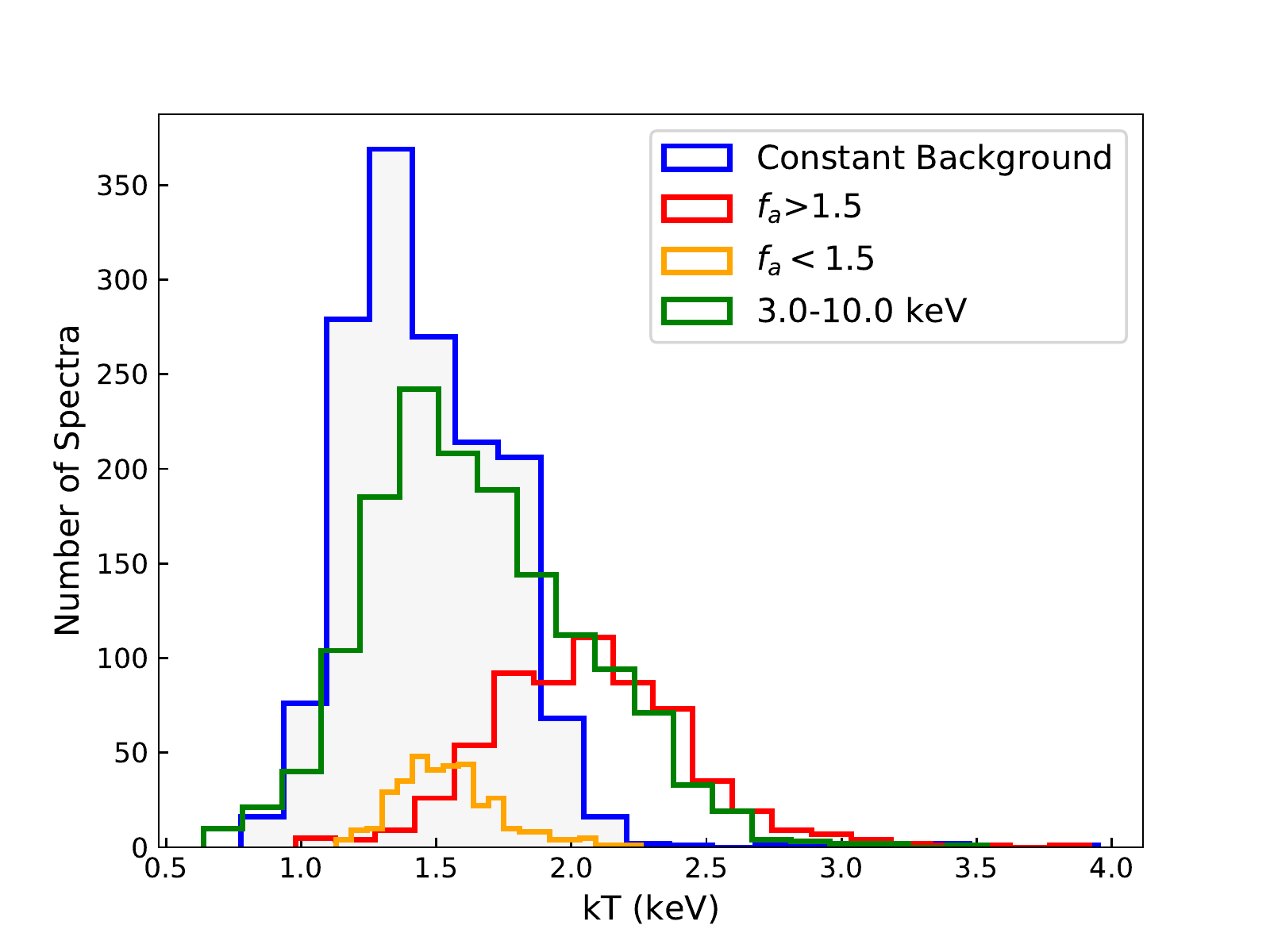}
    \caption{Histogram of the inferred blackbody temperature values obtained using different methods. Use of the \fa parameter forces the blackbody temperature to higher values, especially if its value is large. Note that due to difference in sizes of the samples the number of bins are kept different.}
    \label{fig:comp_hist}
\end{figure}

In total we analyzed 2611 X-ray spectra extracted from 40 thermonuclear X-ray bursts observed with \nicer. Time evolution of the spectral parameters for some of the example bursts are shown in \autoref{fig:bursts_plts_faint} and \autoref{fig:astrosat_bursts_plts} for the bursts simultaneously observed with AstroSat. Of all 2611 spectra, 1512 instances were above a flux limit of $10^{-9}$\fluxcgs\ and of these in 948 cases the use of a scaling factor (\fa) greater than one, is statistically preferred. The flux limit is selected to avoid the scatter in parameters when the burst emission becomes comparable to the persistent state flux. These results indicate that in about 63\% of the X-ray spectra we need a scaling factor to the persistent emission. A histogram of the resulting $\chi^2$/dof values using different approaches is presented in \autoref{fig:chi_hist}, which shows the improvement in the fits. An example on the observed variation in the pre-burst emission with the addition of the \fa parameter is shown in \autoref{fig:spec_var} for both a \nicer~ only case and a \nicer + LAXPC case. Histogram of all the \fa values are shown in \autoref{fig:fa_hist}. We show burst 5, 6, and 15 in different colors in the plot as they show evidence for photospheric radius expansion \citep[following the definitions of][]{Galloway2008,2012ApJ...747...77G}. We note that although required to better fit the data, the use of the scaling factor significantly alters the inferred best fit spectral parameters, which impedes the observation of the expected spectral evolution
from a photospheric radius expansion event, especially regarding the expected evolution of the blackbody normalization. As was the case for Aql~X-1 \citep{2022MNRAS.510.1577G}, the largest \fa values are seen during the peak flux moments of these bursts. Large \fa values also have more significant effects on the inferred spectral parameters for the bursts. The change in the inferred blackbody temperature is shown in \autoref{fig:comp_hist} for constant background approach (for 0.5--10.0~keV and 3--10.0~keV ranges) as well as different \fa values.

\begin{figure}
    \centering
    \includegraphics[scale=0.5]{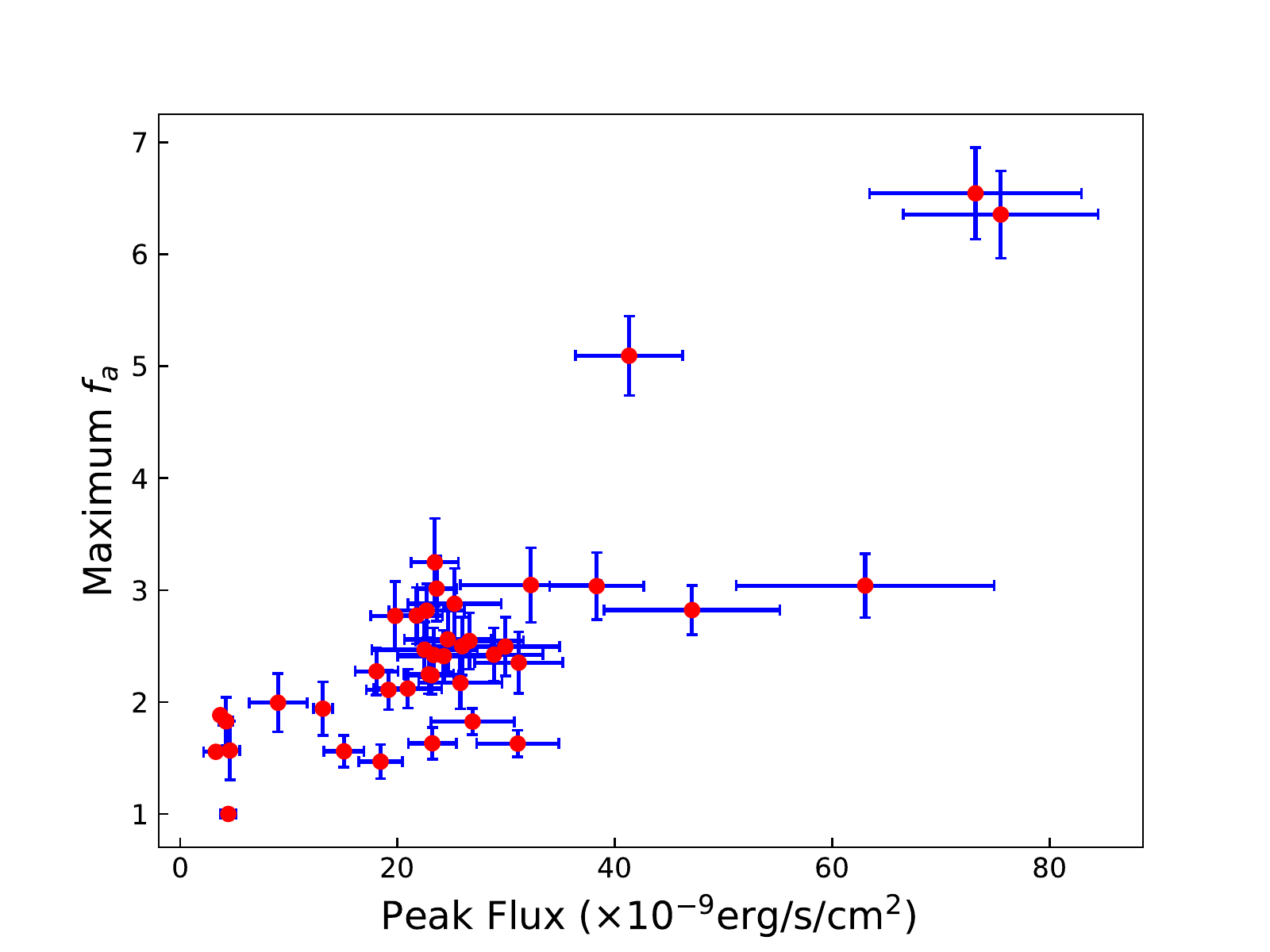}
    \caption{Maximum \fa value reached during a burst is shown as a function of the peak flux.}
    \label{fig:peak_fl_fa}
\end{figure}

We searched for correlations between several parameters and the value of the scaling factor using Pearson correlation. The correlation coefficient between peak flux and the \fa value obtained at that moment  is 0.80, with a chance probability of 7.81$\times10^{-10}$. On the other hand, the correlation coefficient between the peak flux of a burst and maximum \fa value that is reached in that particular burst is 0.827, with a chance probability of 4.68$\times10^{-11}$, indicating a slightly stronger correlation between the peak flux and maximum \fa value. This relation is shown in \autoref{fig:peak_fl_fa}.

\begin{figure}
    \centering
    \includegraphics[scale=0.5]{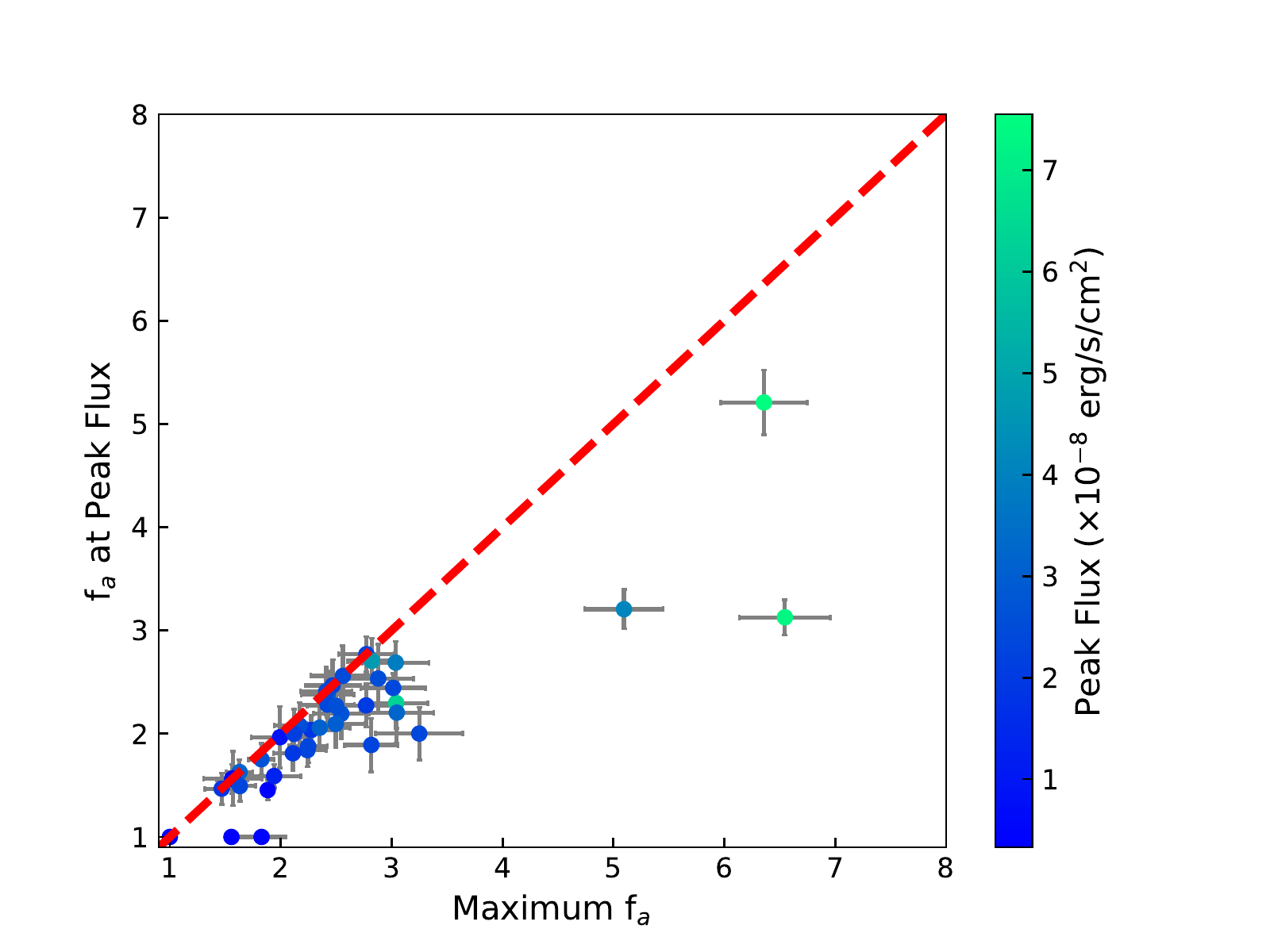}
    \caption{Comparison of the \fa values obtained at the peak flux moment and the maximum \fa value reached during a burst.}
    \label{fig:fa_comp_sp}
\end{figure}

Although in general \fa values obtained at the peak flux moment and the maximum \fa value of a burst agree with each other the correlation between these parameters is not perfect (see \autoref{fig:fa_comp_sp}). Correlation between the maximum \fa value and the \fa at the peak flux moment is 0.86, with a chance probability of  1.25$\times10^{-12}$. The correlation coefficient between persistent flux and the maximum \fa value is only 0.2 with a chance probability of 0.20 indicating no significant correlation between these two parameters. Such lack of correlation allows us to conclude that the increase in the mass accretion rate due to Poynting-Robertson drag, which is characterized here by the \fa parameter, does not seem to depend on the pre-burst spectral state of the source.

\begin{deluxetable*}{ccccccccc}
\label{tab:time_res_results}
\tablehead{
\colhead{BID}&\colhead{Peak \fa}&\multicolumn2c{Peak Flux\tablenotemark{a}}&\multicolumn2c{Peak kT}&\multicolumn2c{Peak Norm}&Fluence\tablenotemark{b}\\
&&\colhead{with \fa}&\colhead{without \fa}&\colhead{with \fa}&\colhead{without \fa}&\colhead{with \fa}&\colhead{without \fa}&with \fa}
\tablecaption{Best fit spectral parameters of the bursts at the peak flux moment.}
\startdata
2 &1.6$\pm$0.1 & 3.1$\pm$0.4&2.6$\pm$0.2 &2.2$\pm$0.2 & 1.7$\pm$0.1 & 123$\pm$28&270$\pm$34&12.3$\pm$0.4\\
3 &1.6$\pm$0.1 & 1.5$\pm$0.2 & 1.5$\pm$0.1 & 1.9$\pm$0.2 &1.5$\pm$0.1&108$\pm$25& 303$\pm$38&7.8$\pm$0.3\\
5 &3.2$\pm$0.2 & 4.1$\pm$0.5 & 3.0$\pm$0.2 & 2.4$\pm$0.2 & 1.4$\pm$0.1 & 117$\pm$24&699$\pm$71&28.1$\pm$0.7\\
6 &3.1$\pm$0.2 & 7.3$\pm$1.0 & 4.6$\pm$0.3 & 3.1$\pm$0.3 & 1.3$\pm$0.1 & 79$\pm$15&1422$\pm$148&37.9$\pm$1.0\\
7 & -- & 0.4$\pm$0.1 & 0.4$\pm$0.1 & 1.4$\pm$0.2 & 1.4$\pm$0.2&100$\pm$29 &100$\pm$29&1.7$\pm$0.1\\
9 &2.3$\pm$0.3 & 6.3$\pm$1.2 & 4.4$\pm$0.7 & 3.9$\pm$0.6&2.8$\pm$0.3&33$\pm$9&68$\pm$12&31.5$\pm$1.2\\
11 &2.7$\pm$0.2&3.8$\pm$0.4&2.8$\pm$0.2&2.8$\pm$0.2&2.1$\pm$0.1&65$\pm$10& 141$\pm$14&21.7$\pm$1.0\\
12 &2.6$\pm$0.3&2.46$\pm$0.4& 1.8$\pm$0.2&2.4$\pm$0.3& 1.8$\pm$0.1& 76$\pm$19&173$\pm$29&8.9$\pm$0.4\\
13 &1.6$\pm$0.3&0.5$\pm$0.1&0.5$\pm$0.1&1.6$\pm$0.2&1.4$\pm$0.1&73$\pm$28&125$\pm$29&1.6$\pm$0.1\\
14 &2.8$\pm$0.2&2.2$\pm$0.3&1.7$\pm$0.2&2.3$\pm$0.2&1.6$\pm$0.1&78$\pm$14&231$\pm$32&16.3$\pm$0.6\\
15 &5.2$\pm$ 0.3&7.6$\pm$ 0.9 & 5.5$\pm$ 0.4&2.3$\pm$0.2& 1.2$\pm$ 0.1&247$\pm$52&2374$\pm$265&25.1$\pm$1.1\\
16 & 1.5$\pm$0.2&2.3$\pm$0.2&2.2$\pm$0.2&1.9$\pm$0.1& 1.7$\pm$0.1&176$\pm$33&270$\pm$32& 15.0$\pm$0.4\\
18 &1.5$\pm$ 0.2&1.8$\pm$0.2& 1.7$\pm$0.2& 1.9$\pm$0.1& 1.7$\pm$0.1& 128$\pm$26& 190$\pm$27&8.5$\pm$0.3 \\
19 &2.0$\pm$0.1&1.8$\pm$0.2& 1.6$\pm$0.2& 2.3$\pm$0.2 &1.7$\pm$0.1 &63$\pm$11&177$\pm$25&16.9$\pm$0.7\\
20 &1.8$\pm$ 0.2& 2.7$\pm$0.4& 2.2$\pm$0.2& 2.4$\pm$0.2& 1.9$\pm$0.1& 85$\pm$19& 165$\pm$25&17.3$\pm$0.6\\
21 &2.4$\pm$0.1&2.4$\pm$0.2&1.9$\pm$0.1&2.3$\pm$0.1&1.8$\pm$0.1&76$\pm$9&158$\pm$12&19.2$\pm$0.7\\
22 & 1.8$\pm$0.2& 2.3$\pm$0.3& 2.0$\pm$0.2& 2.3$\pm$0.2& 1.9$\pm$0.1&83$\pm$14&147$\pm$21&16.8$\pm$0.7\\
23 &1.9$\pm$0.2&2.3$\pm$0.2& 2.0$\pm$0.2& 2.3$\pm$0.1& 1.9$\pm$0.1&82$\pm$13&139$\pm$15&18.3$\pm$0.7\\
24 &2.7$\pm$0.2&4.7$\pm$0.8& 3.1$\pm$0.4&3.3$\pm$0.4&2.2$\pm$0.2&437$\pm$11& 120$\pm$18&25.7$\pm$1.1\\
25 & -- &0.4$\pm$0.1& 0.4$\pm$0.1&1.5$\pm$0.1&1.5$\pm$0.1&87$\pm$22&87$\pm$22&15.0$\pm$0.2\\
26 &2.1$\pm$0.2& 2.6$\pm$0.4& 2.1$\pm$0.3&2.5$\pm$0.2&2.0$\pm$0.1&68$\pm$15&127$\pm$19&21.3$\pm$0.7\\
27 &2.4$\pm$0.2&2.9$\pm$0.5&2.2$\pm$0.3&2.6$\pm$0.3&2.0$\pm$0.1&60$\pm$14&139$\pm$21&19.4$\pm$0.7\\
28 &2.3$\pm$0.2& 2.3$\pm$0.3&1.9$\pm$0.2&2.4$\pm$0.2&1.9$\pm$0.1&73$\pm$17&150$\pm$22&19.5$\pm$0.7\\
29 &2.0$\pm$0.3& 2.3$\pm$0.2&2.2$\pm$0.2&2.3$\pm$0.1&2.0$\pm$0.1&86$\pm$14& 131$\pm$13& 19.1$\pm$0.9\\
30 &2.5$\pm$0.3& 2.5$\pm$0.4&2.2$\pm$0.3& 2.6$\pm$0.3& 2.1$\pm$0.1& 55$\pm$15& 110$\pm$16&23.5$\pm$0.9\\
32 &2.2$\pm$0.3 & 3.2$\pm$ 0.7 & 2.5$\pm$0.4& 3.2$\pm$0.5& 2.4$\pm$0.2& 35$\pm$11& 75$\pm$13&24.7$\pm$1.0\\
34 &2.3$\pm$0.2&2.0$\pm$0.2& 1.8$\pm$0.2& 2.4$\pm$0.2& 1.9$\pm$0.1& 58$\pm$11& 145$\pm$22& 18.6$\pm$0.9\\
35 &2.0$\pm$0.3 & 0.9$\pm$0.3& 0.8$\pm$0.1& 2.3$\pm$0.1& 1.7$\pm$0.2& 32$\pm$15& 86$\pm$19&4.2$\pm$0.3\\
36 &2.1$\pm$0.2& 3.1$\pm$0.4 & 2.5$\pm$0.3& 2.9$\pm$0.3& 2.3$\pm$0.1& 45$\pm$9& 86$\pm$10& 18.5$\pm$1.0\\
38 &2.2$\pm$0.2&2.7$\pm$0.5&2.0$\pm$0.3& 2.7$\pm$0.3&2.0$\pm$0.2&50$\pm$14&120$\pm$20&21.5$\pm$0.7\\
39 & 2.5$\pm$0.3&2.3$\pm$0.5&1.9$\pm$0.2&2.8$\pm$0.4&1.9$\pm$0.1&39$\pm$13&150$\pm$20&23.1$\pm$0.8\\
40 &1.5$\pm$0.1 & 0.4$\pm$ 0.1 & 0.4$\pm$0.1 & 1.9$\pm$0.2 & 1.5$\pm$0.1 & 30$\pm$9&70$\pm$11&1.9$\pm$0.2\\
41 &1.9$\pm$0.3 &2.3$\pm$ 0.4 & 2.0$\pm$0.2 & 2.4$\pm$0.3 & 2.0$\pm$0.1 &64$\pm$16& 111$\pm$17&20.1$\pm$0.7\\
42 &2.0$\pm$0.2 &2.1$\pm$ 0.3 & 1.9$\pm$0.2& 2.3$\pm$0.2 & 1.9$\pm$0.1 & 69$\pm$17& 130$\pm$20& 9.3$\pm$0.5\\
43 & -- & 0.3$\pm$0.1 & 0.3$\pm$0.1 & 2.0$\pm$0.2 & 2.1$\pm$0.2 & 21$\pm$10& 21$\pm$10& 1.9$\pm$0.3\\
45 &2.4$\pm$0.2&2.4$\pm$0.4 & 2.0$\pm$0.2 & 2.7$\pm$0.3 & 2.1$\pm$0.1&50$\pm$14&103$\pm$16&23.7$\pm$0.8 \\
46 &1.6$\pm$0.1&1.3$\pm$0.1 & 1.2$\pm$0.1 & 1.9$\pm$0.1 & 1.7$\pm$0.1 & 92$\pm$11& 140$\pm$11&5.5$\pm$0.3 \\
47 &1.8$\pm$0.2 & 1.9$\pm$0.2 & 1.7$\pm$0.1 & 2.3$\pm$0.2 & 1.9$\pm$0.1 &67$\pm$12& 117$\pm$13& 15.3$\pm$0.7\\
50 &2.1$\pm$0.2 & 3.0$\pm$0.5 & 2.4$\pm$0.3 & 2.8$\pm$0.3 & 2.2$\pm$0.2 &49$\pm$12&100$\pm$16& 24.8$\pm$0.9\\
51 &2.3$\pm$0.2 & 2.6$\pm$0.3 & 2.0$\pm$0.2 & 2.8$\pm$0.2 & 2.0$\pm$0.1 &45$\pm$9&112$\pm$13& 17.9$\pm$0.8\\
\enddata
    \tablenotetext{a}{Unabsorbed bolometric flux in units of $\times10^{-8}$ \fluxcgs.}
    \tablenotetext{b}{Fluences are in units of $\times10^{-8}$~\flencecgs.}
\end{deluxetable*}    


\begin{figure}
    \centering
    \includegraphics[scale=0.5]{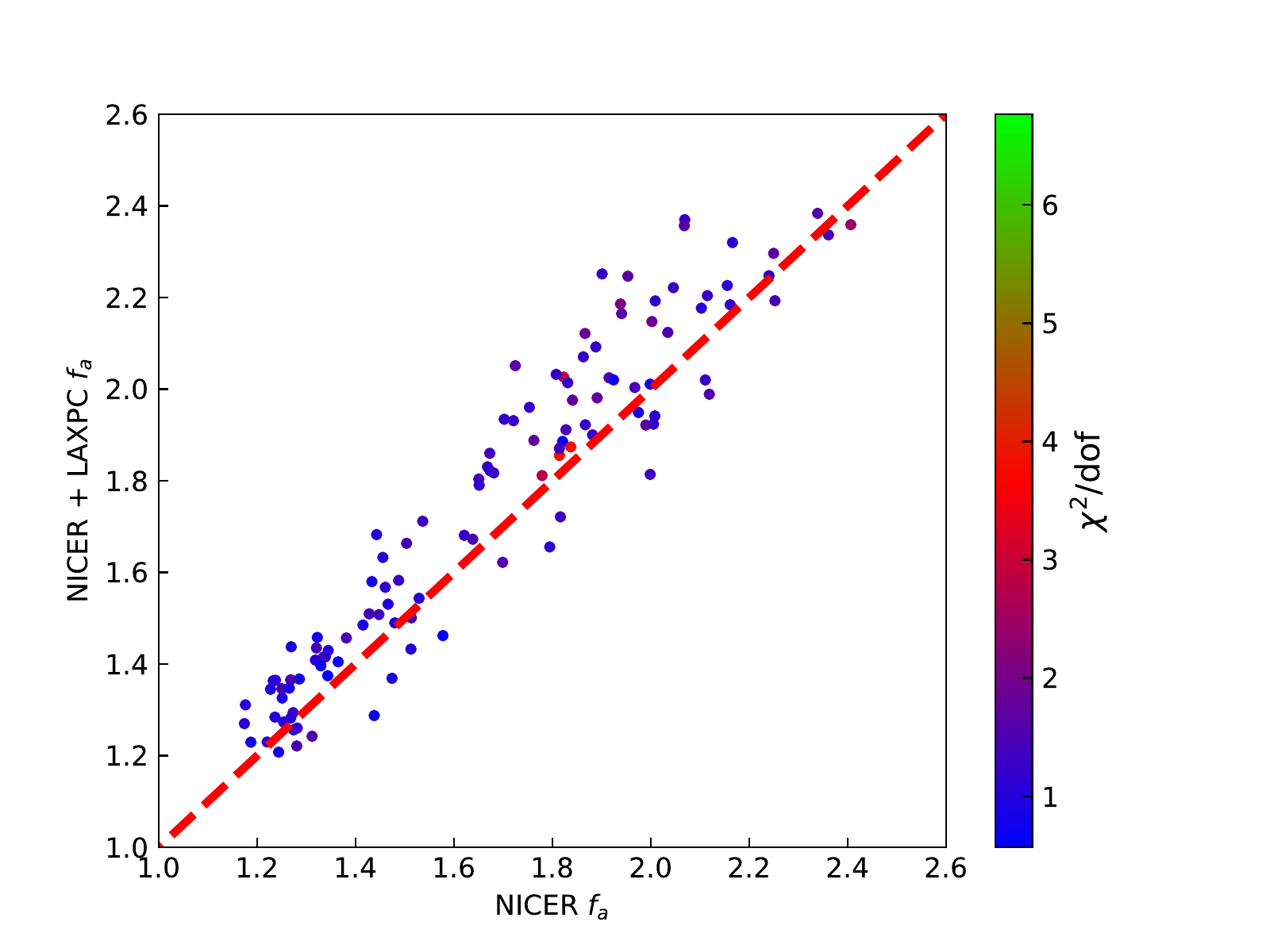}
    \includegraphics[scale=0.5]{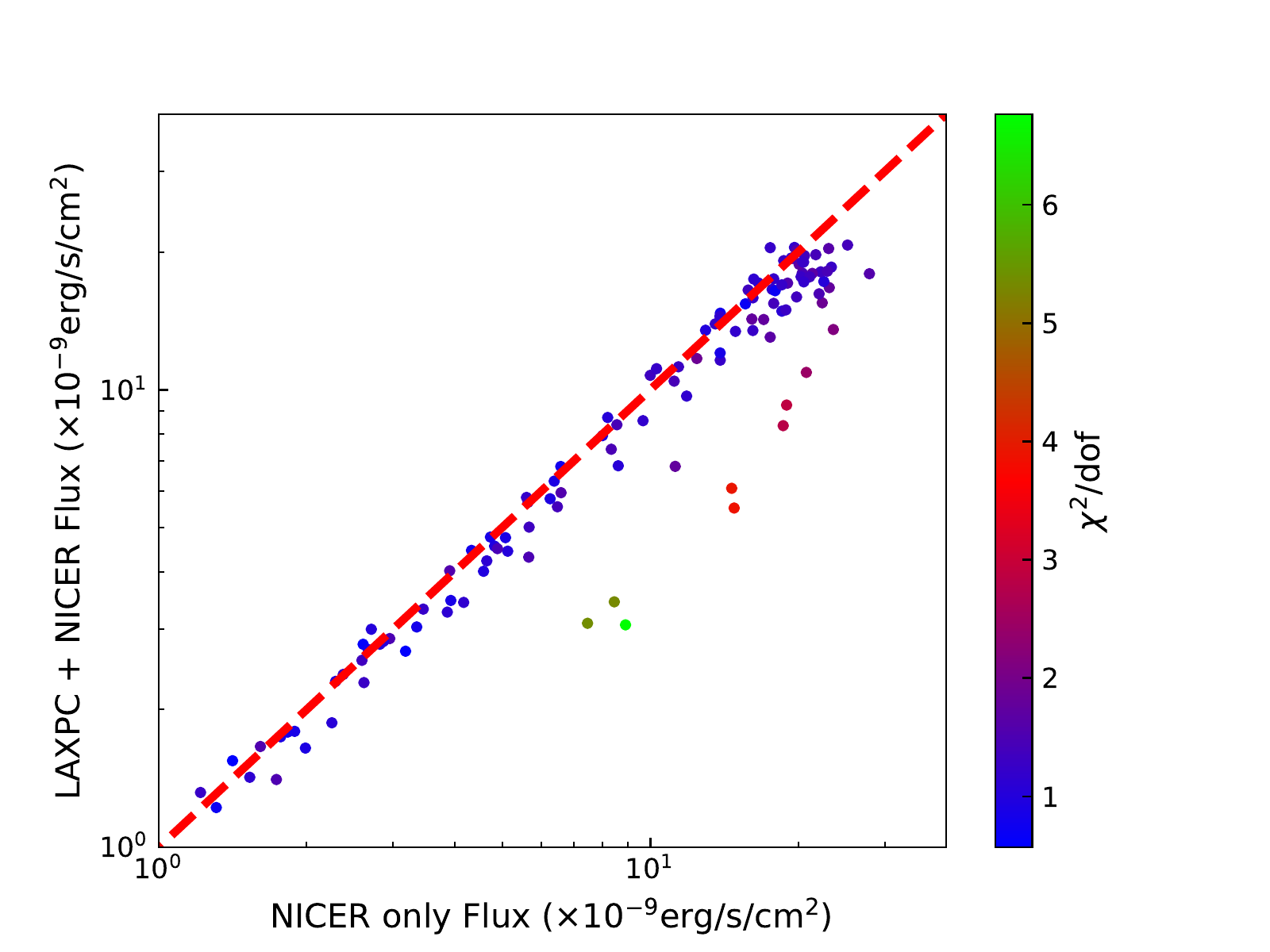}
    \includegraphics[scale=0.5]{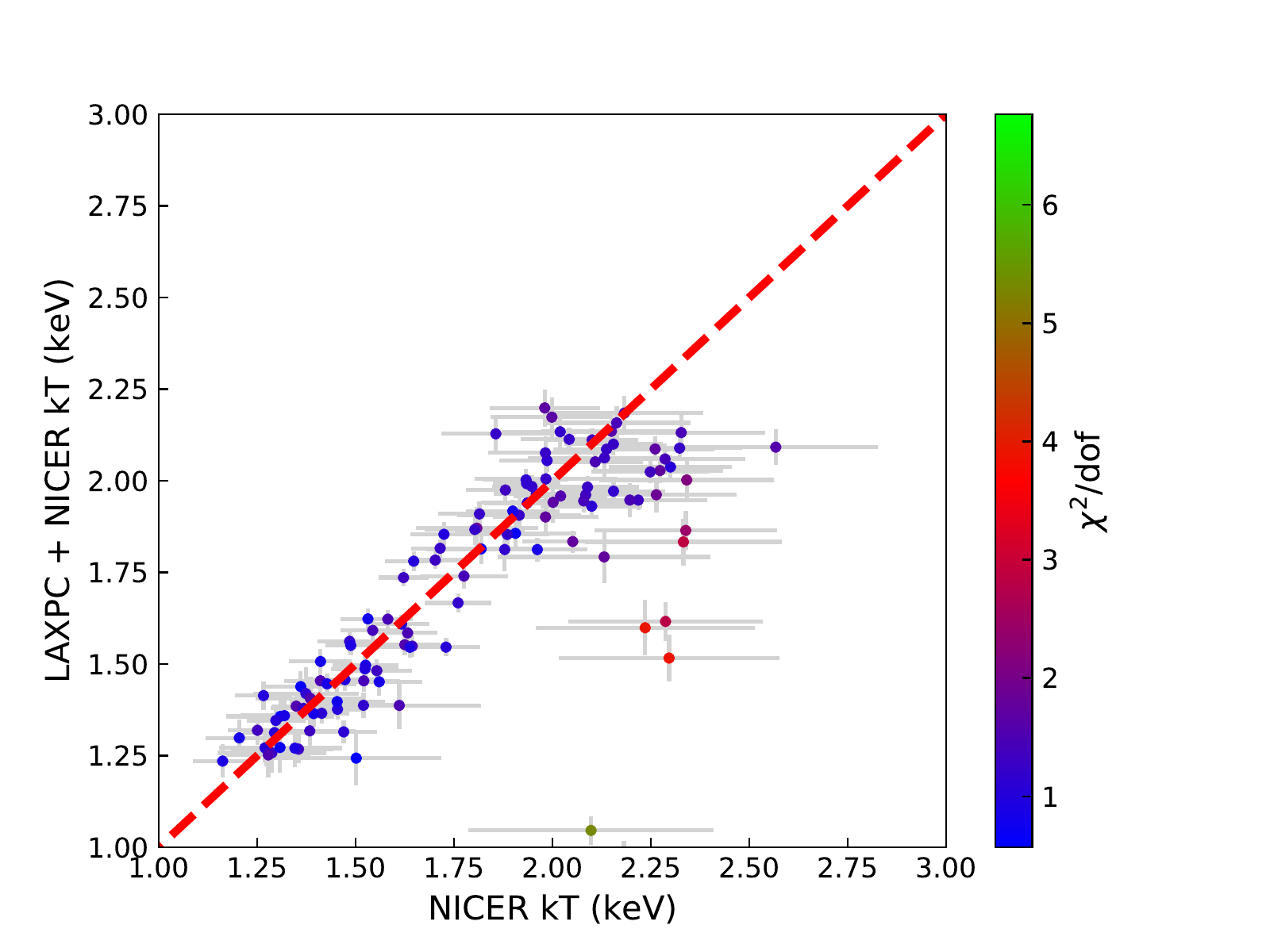}
    \caption{Comparison of the best fit \fa, flux, and kT values obtained when using only \nicer~ and \nicer~ together with LAXPC. The lowest panel also shows the uncertainties of the best fit kT values in both axis, to emphasize the decrease in the measurement uncertainty with the addition of the LAXPC data.}
    \label{fig:fa_comp_laxpc}
\end{figure}

Results of the fits obtained for five bursts where we had simultaneous data from both \nicer~and AstroSat/LAXPC were similar to the results obtained when we used only the \nicer~data. A comparison of the best fit \fa, blackbody temperature, and bolometric flux values are shown in \autoref{fig:fa_comp_laxpc}. We find that the addition of LAXPC data does not significantly alter the best-fit values of the spectral analysis, although it does improve the measurement uncertainties. For example the uncertainties in the burst temperature values decrease on average by a factor of three especially at higher temperatures (see lower panel of \autoref{fig:fa_comp_laxpc}). While a few spectra do yield different outcomes in the \nicer~ data alone as compared to NICER+LAXPC analysis (as indicated by the offset from the dashed line), those cases always have poor $\chi^{2}$/dof values. We note that simultaneous fits generally result in slightly lower \fa values compared to just \nicer~results.

\subsection{Application of a reflection model}

\begin{figure}
    \centering
    \includegraphics[scale=0.5]{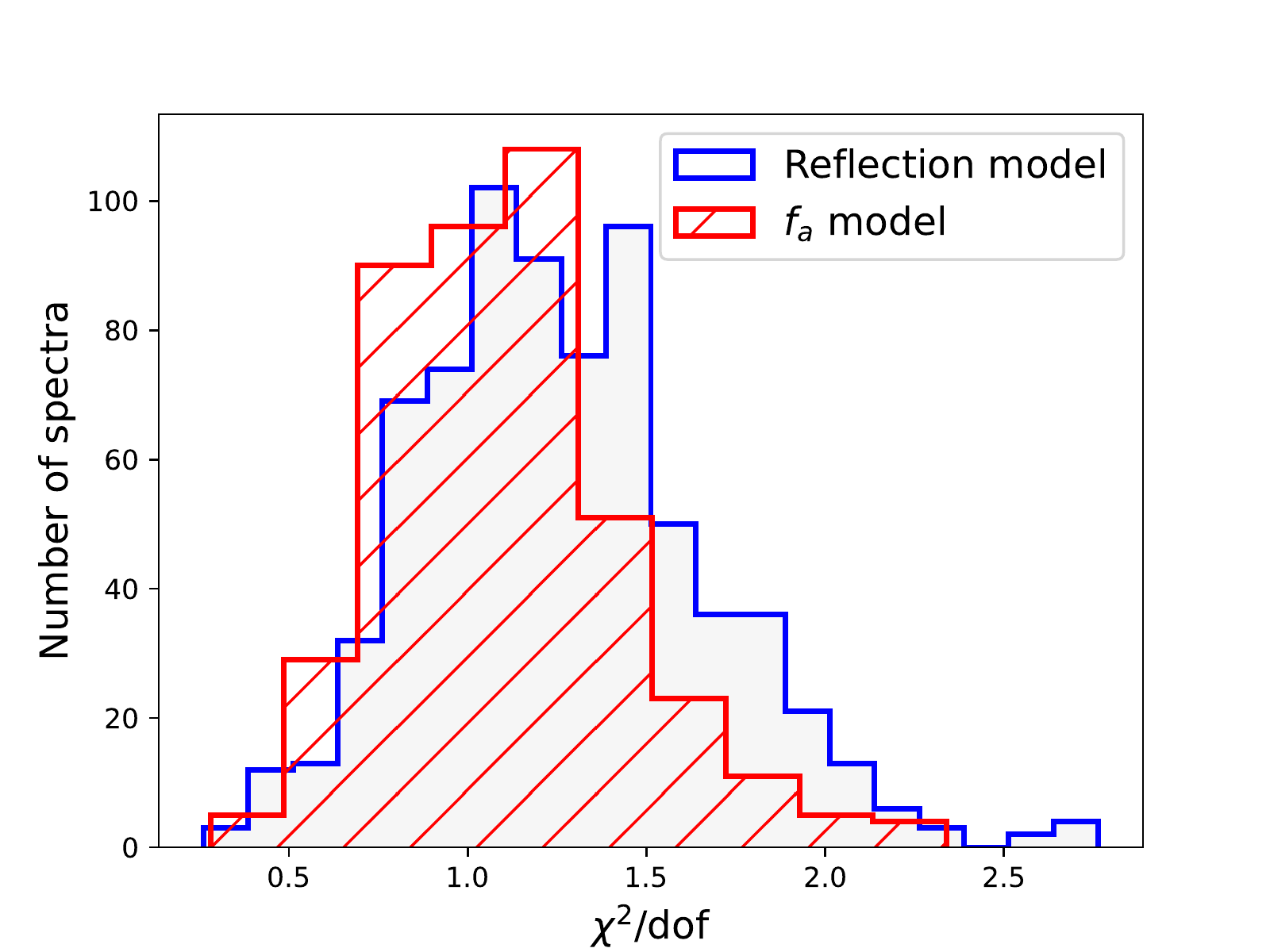}
    \caption{Comparison of $\chi^2$/dof values obtained when using the \fa approach or the reflection model of \nicer~bursts.}
    \label{fig:reflfa_chi_comp}
\end{figure}

We applied the \emph{bbrefl} model to all the X-ray spectra where the \fa was greater than one and the bolometric flux was larger than half the peak flux of a burst. Our results generally agree very well with our previous analysis using \nicer~ observations of bursts from Aql~X-1 \citep{2022MNRAS.510.1577G}. We find that when allowed to vary, the ionization parameter reaches the largest tabulated value. We therefore fixed this parameter in our analysis. This indicate very high ionization levels during the burst in the accretion disk. As a result, the reflection model provides a similar but slightly worse improvement on the fits to what the \fa model provides (see \autoref{fig:reflfa_chi_comp}) for the same degrees of freedom.  

\begin{figure}
    \centering
    \includegraphics[scale=0.5]{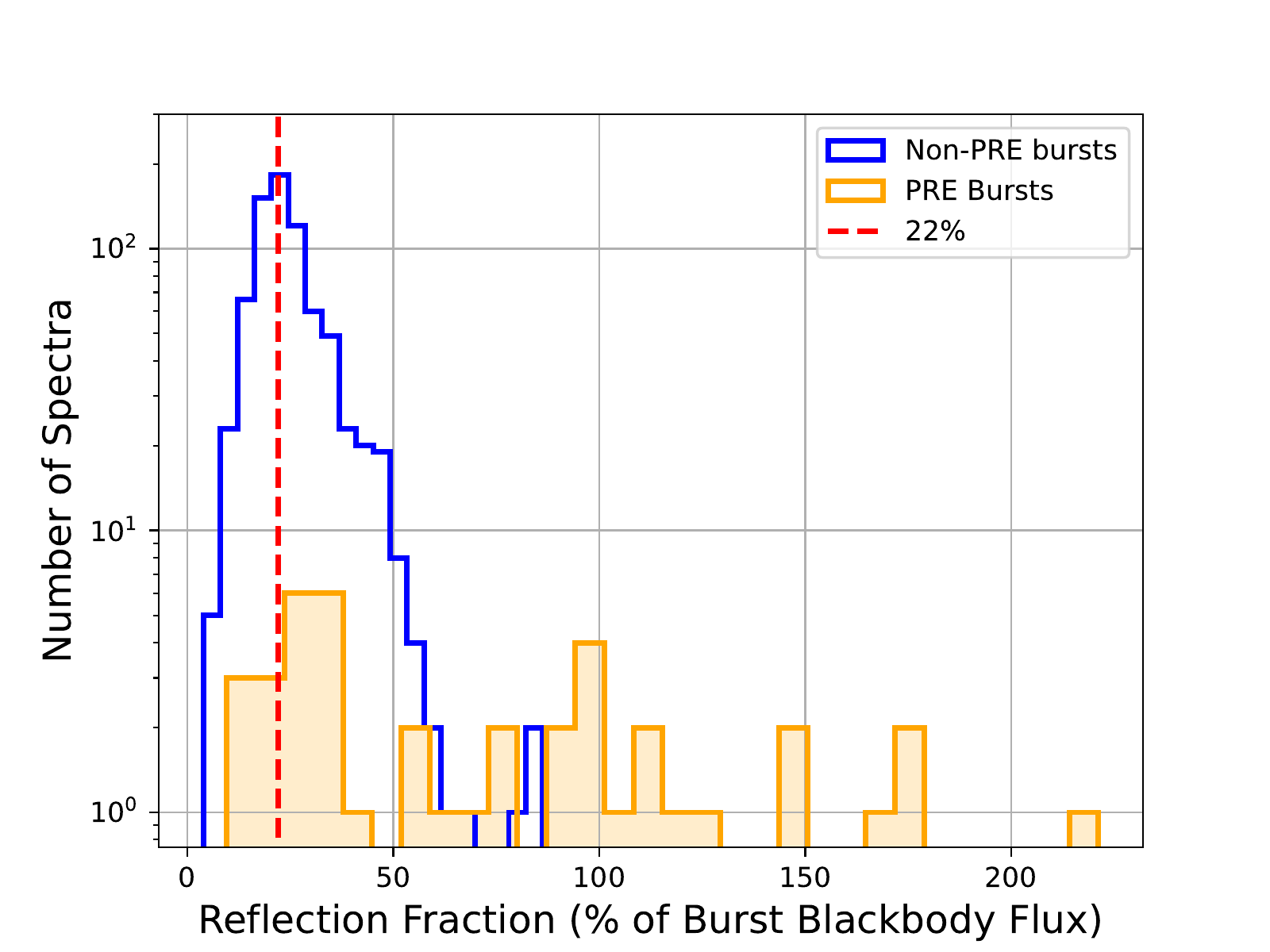}
    \includegraphics[scale=0.5]{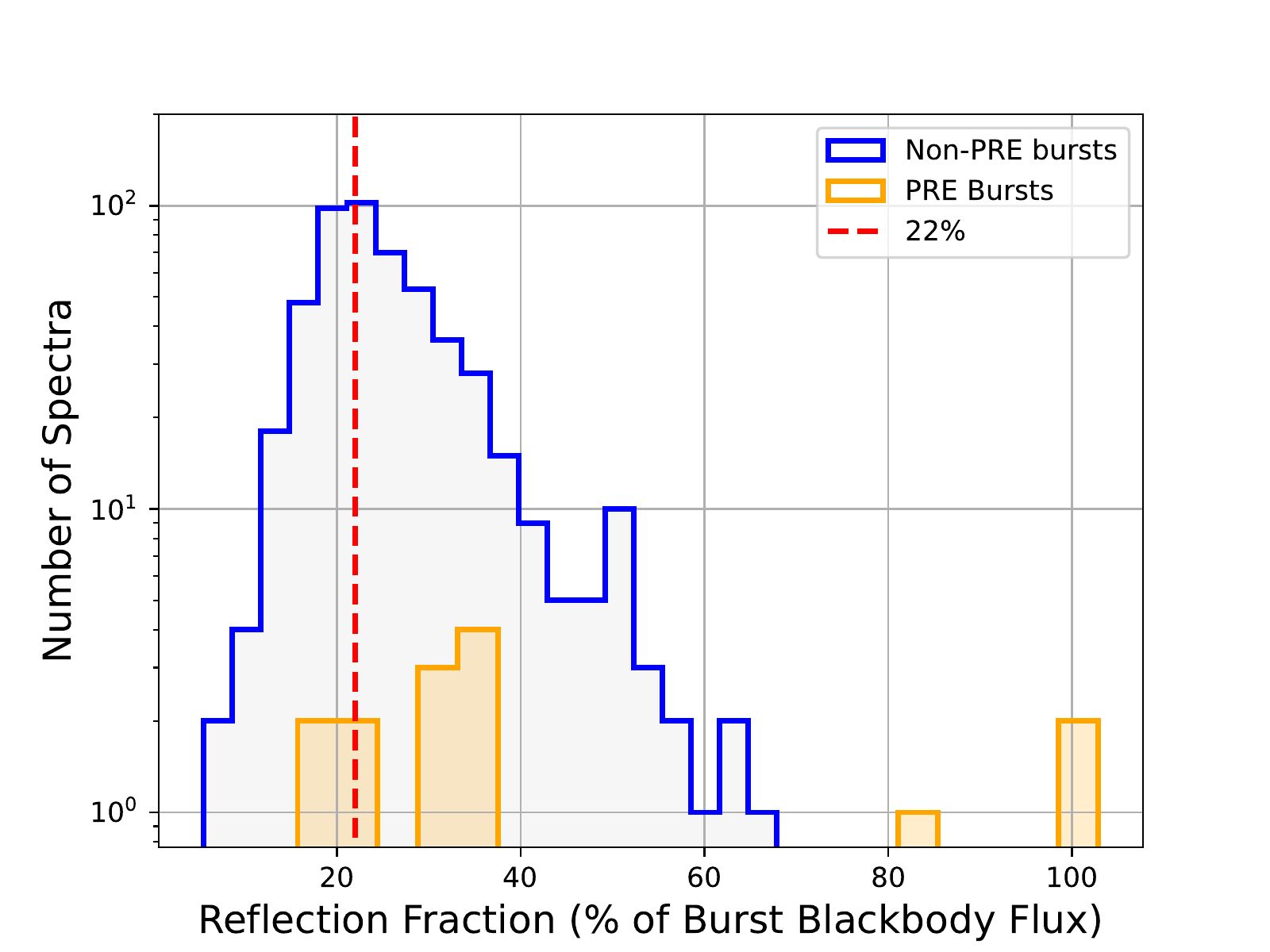}
    \caption{Histogram of the ratio of the flux of the reflection model and  the flux of the corresponding the burst blackbody. 
    Upper panel shows the case when all of the X-ray spectra are included, while the lower panel shows the results when the fit statistics is acceptable 
    (as defined by the range shown in \autoref{fig:reflection_chi2}).
    In orange, results from the bursts that show evidence of a photospheric radius expansion are shown 
    (as shown in \autoref{tab:burst_prop}). In both panels vertical red dashed lines show the 22\% ratio.}
    \label{fig:refl_frac}
\end{figure}

\begin{figure}
    \centering
    \includegraphics[scale=0.5]{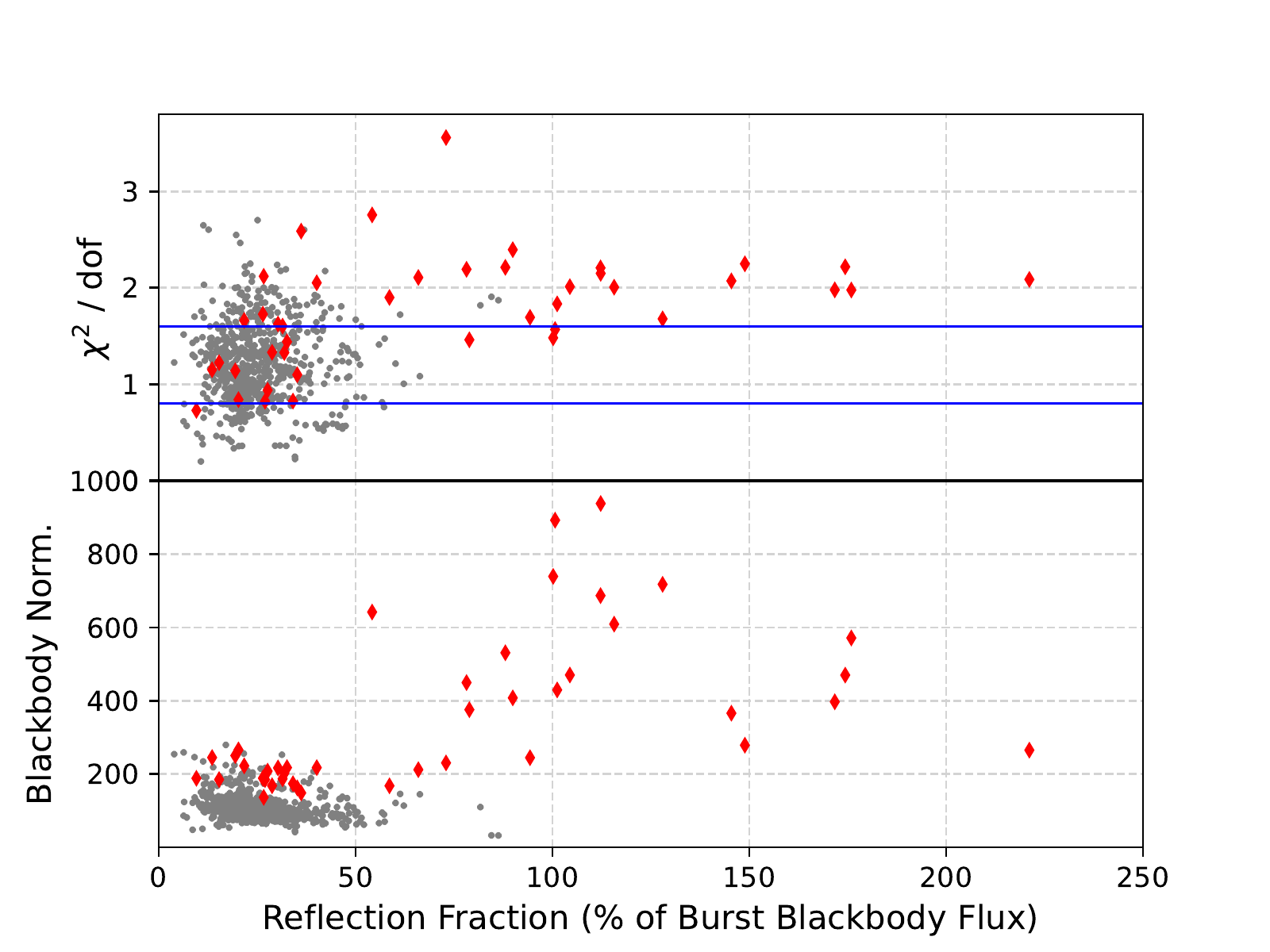}
    \caption{{\it Upper Panel:} The fit statistics as a function of the ratio of the flux of the reflection model and the flux of the corresponding blackbody model. Red diamonds show the results for bursts that show evidence of a photospheric radius expansion 
    (as given in \autoref{tab:burst_prop}), while the grey dots show the values for the rest of the bursts. 
    The blue horizontal solid lines show the range where the fit is assumed to be statistically acceptable. {\it Lower Panel:} Blackbody normalization~($\rm{R}^{2}_{km}/\rm{D}^{2}_{10kpc}$) as a function of reflection fraction. It can be seen that the reflection fraction tends to be larger when the blackbody normalization is larger.} 
    \label{fig:reflection_chi2}
\end{figure}

We calculated the fraction of the unabsorbed flux of the reflection model to that of the burst blackbody in the 0.5--10.0~keV range. We find that for the great majority of the bursts, reflection fraction is about 22\% (see \autoref{fig:refl_frac}). Only for the bursts that show photospheric radius expansion, the reflection fraction can be larger than the incident photon flux by a factor of two. When the reflection fraction is plotted against the fit statistics (upper panel of \autoref{fig:reflection_chi2}) and against the normalization of the blackbody model (lower panel of \autoref{fig:reflection_chi2}) it can be seen that the reflection fraction is very large only when the fit statistics are significantly bad and these cases are seen when the blackbody normalization is significantly larger indicating that it is seen when the photosphere expanded. Again similar to our results from Aql~X-1, we see that especially during the photospheric radius expansion phases of the bursts reflection model by itself can not explain all of the excess emission in the X-ray spectrum.  If we only take the results obtained when the $\chi^2$/dof is within an acceptable range  (0.8 $<$ $\chi^2$/dof $<$ 1.6) then the resulting distribution of reflection fractions peak again at around 22\% with almost no result exceeding the seed photon flux. This tight clustering is expected and supports the reflection scenario since from burst to burst the emitting surface area, which would be the full surface of the neutron star and the reflecting surface, which would be the inner accretion disk can be assumed to be constant.

\begin{figure}
    \centering
    \includegraphics[scale=0.5]{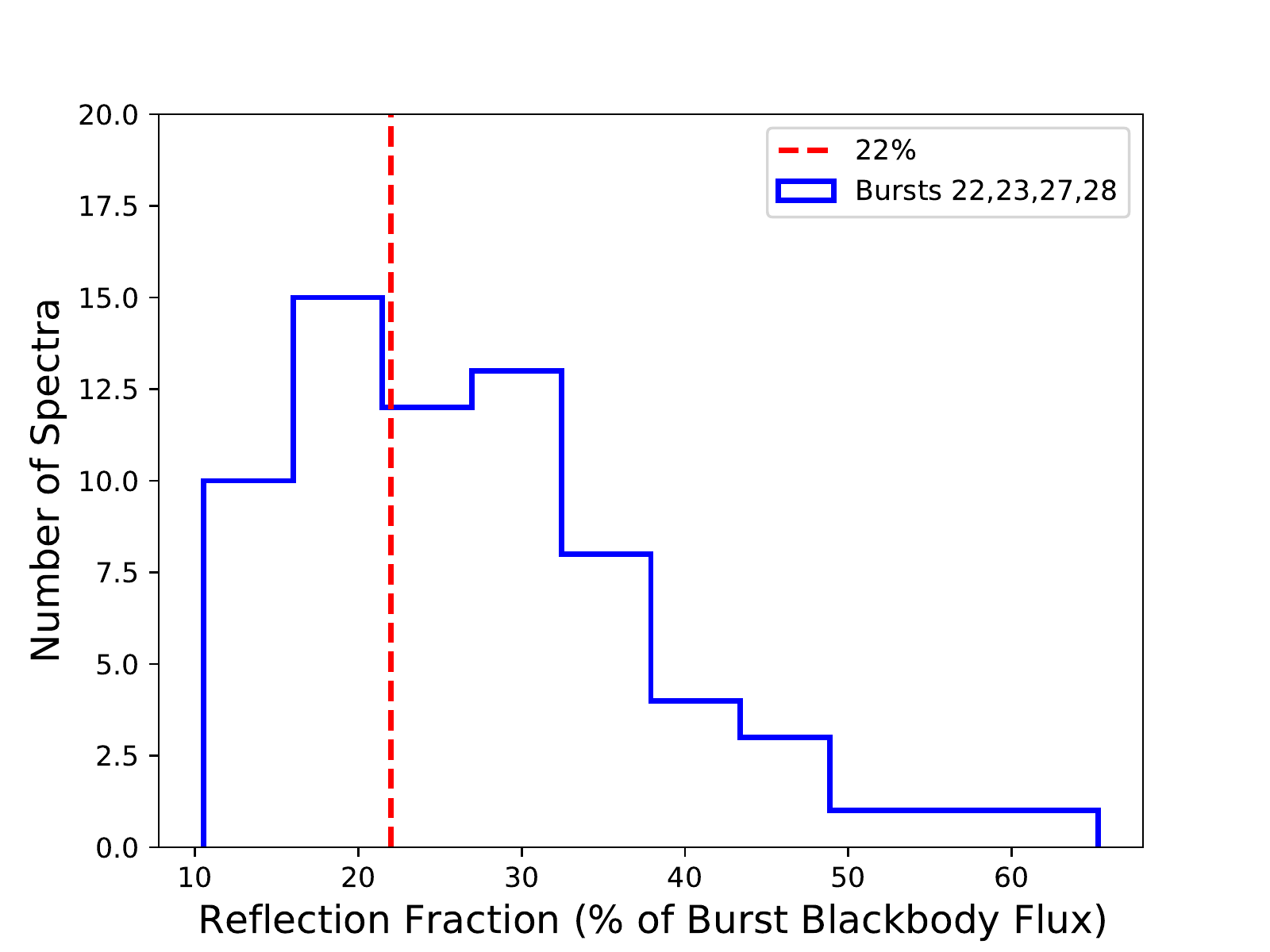}
    \caption{Histogram of the ratio of the flux of the reflection model and  the flux of the corresponding burst blackbody for the X-ray spectra extracted from bursts 22, 23, 27, and 28 when the NICER and AstroSat/LAXPC data simultaneously fit. Vertical red dashed lines show the 22\% ratio.}
    \label{fig:refl_frac_astro}
\end{figure}

We also applied the same reflection model to bursts 22, 23, 27, and 28 which are simultaneously observed with \nicer~ and AstroSat. Our simultaneous fit results generally agree very well with the results obtained from bursts with only \nicer~ data. The distribution of the reflection fraction for these bursts are shown in \autoref{fig:refl_frac_astro}. These bursts are generally dimmer compared to more brighter bursts in the \nicer~ sample and the inferred reflection fractions from simultaneous fits follow the distribution for those bursts.

\subsection{Search for evidence of Compton cooling using \nustar~data}

\begin{figure*}
    \centering
    \includegraphics[scale=0.70]{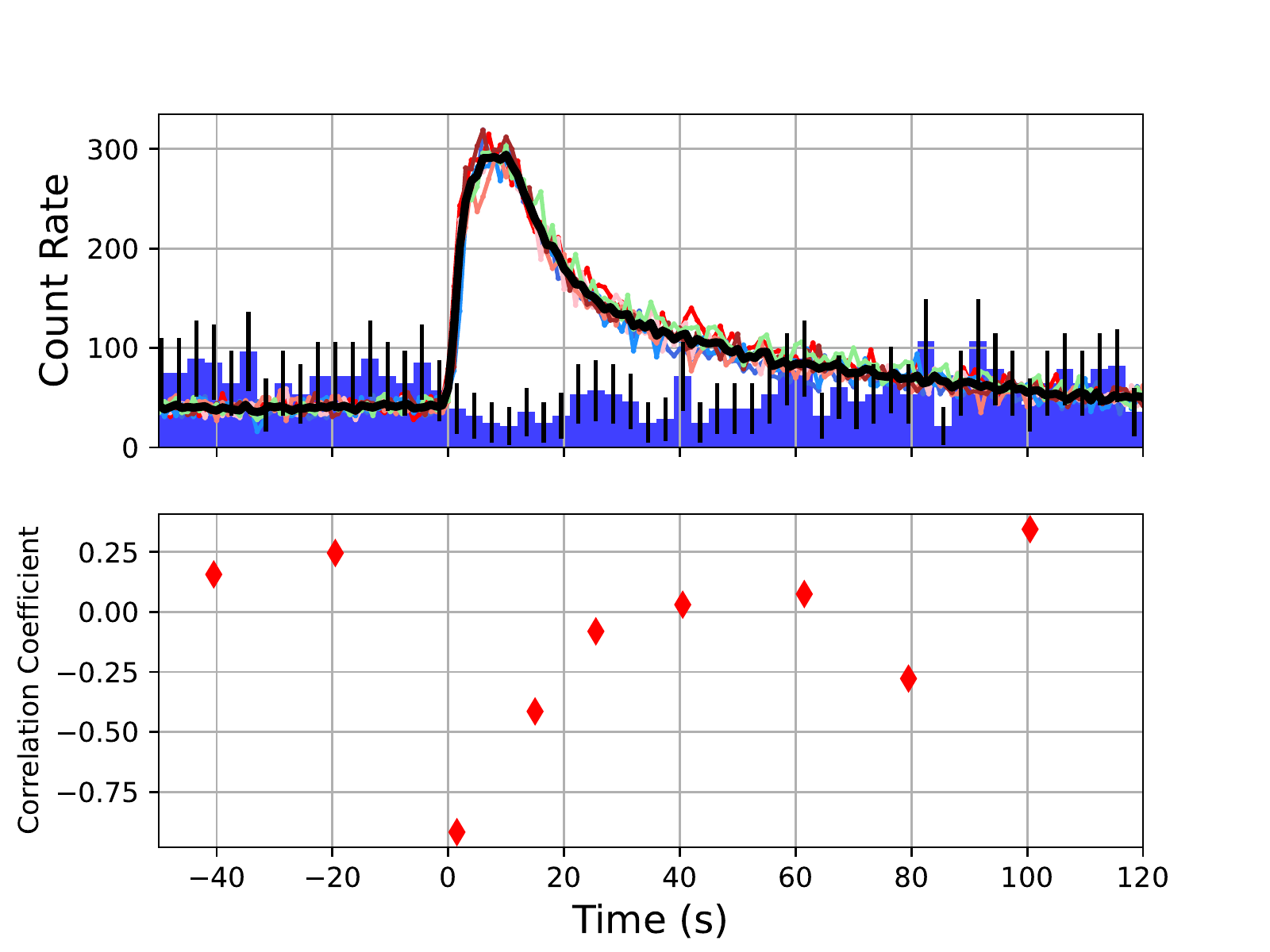}
    \caption{Upper panel shows the lightcurves of some of the X-ray bursts as observed by \nustar~ in the 5--30 and 30--79~keV bands. For the soft band individual bursts in different colors are shown, where the black solid curve shows the average and the time bin size is 1~s. For the hard X-ray band only the average burst lightcurve is shown in blue with appropriate error bars in black. For the hard X-ray lightcurve the time bin size is 4~s and for clarity the number of photons are multiplied by 50. Lower panel, shows the evolution of the pearson correlation coefficient as a function of time in between the two average curves after being rebinned to 5~s and 10~s, in between -5 to +25 seconds since the start of the burst and for the rest, respectively.}
    \label{fig:nustar_burst_lc}
\end{figure*}

It has been suggested \citep{1977Natur.270..502L,1980A&A....86..121S,1982MNRAS.199..763G} and shown \citep{1982Natur.296..226G,1986MNRAS.221..931F,1988ApJ...324..363W} that sudden increase of the number of relatively soft photons due to thermonuclear X-ray bursts can cause the corona surrounding the neutron star to cool down via Compton scattering of these photons with energetic electrons. \cite{2006ApJ...651..416F} has shown that indeed there is observational evidence for a decrease in observed brightness of \source in between 30--79~keV band during burst times by using data from 1 burst observed by INTEGRAL/JEM-X. Studying 114 bursts,  \cite{2013MNRAS.432.2773J} also showed evidence for a shortage in hard X-rays, which is attributed as the observational signature of Compton cooling from \source. A similar analysis has recently been performed using Insight/HXMT data by \cite{2018ApJ...864L..30C}. Having a large dataset we also searched for evidence of Compton cooling in the \nustar~data of \source. As stated in Section~\ref{sec:analysis}, we identified 15 X-ray bursts, six of which were simultaneously observed with \nicer, within the \nustar~observation performed in 2019. The total detected count-rates are relatively lower in \nustar~data we therefore used 7 bursts, which have very similar burst profiles, to calculate an averaged higher signal-to-noise burst profile. Four of these bursts are also simultaneously observed with \nicer~(bursts 44, 45, 46, and 50). We generated two burst profiles in the 3--30~keV and 30--79~keV bands. These bursts and the average burst profiles are shown in \autoref{fig:nustar_burst_lc}. Already from the upper panel of \autoref{fig:nustar_burst_lc} a slight decrease in intensity in the hard X-ray band accompanied by the start of the burst can be seen. Specifically the average count rate we measure in the hard X-ray band 50~s before the bursts is 1.44$\pm$0.07 counts/s, while the average count rate measured in the first 10~s of the bursts is 0.64$\pm$0.07 counts/s. To further investigate whether there is any correlated variation we calculated Pearson correlation coefficient between the soft and hard band lightcurves by taking all the data in 5 and 10~s long intervals in between $-$5~to~$+$25 seconds since the start of the burst and for the rest, respectively. The resulting correlation coefficients are shown in the lower panel of \autoref{fig:nustar_burst_lc}. A strong negative correlation can be seen in the time interval just around the peaks of the bursts which is not visible at other times within roughly the 180~s data around these 7 bursts. We note that since the count-rates are very low in the hard band we did not try to just subtract a background from both soft and hard lightcurves.

\section{Conclusions} 
\label{sec:dicussion}

We report the detection 51 thermonuclear X-ray bursts observed from \source by \nicer~since June 2017. Within this large burst sample we see 2 triple and 4 double burst events, with recurrence times as short as $\sim$400 seconds. One of the triple burst events has already been reported by \cite{2022A&A...660A..31Z}. 

\begin{figure}
    \centering
    \includegraphics[scale=0.5]{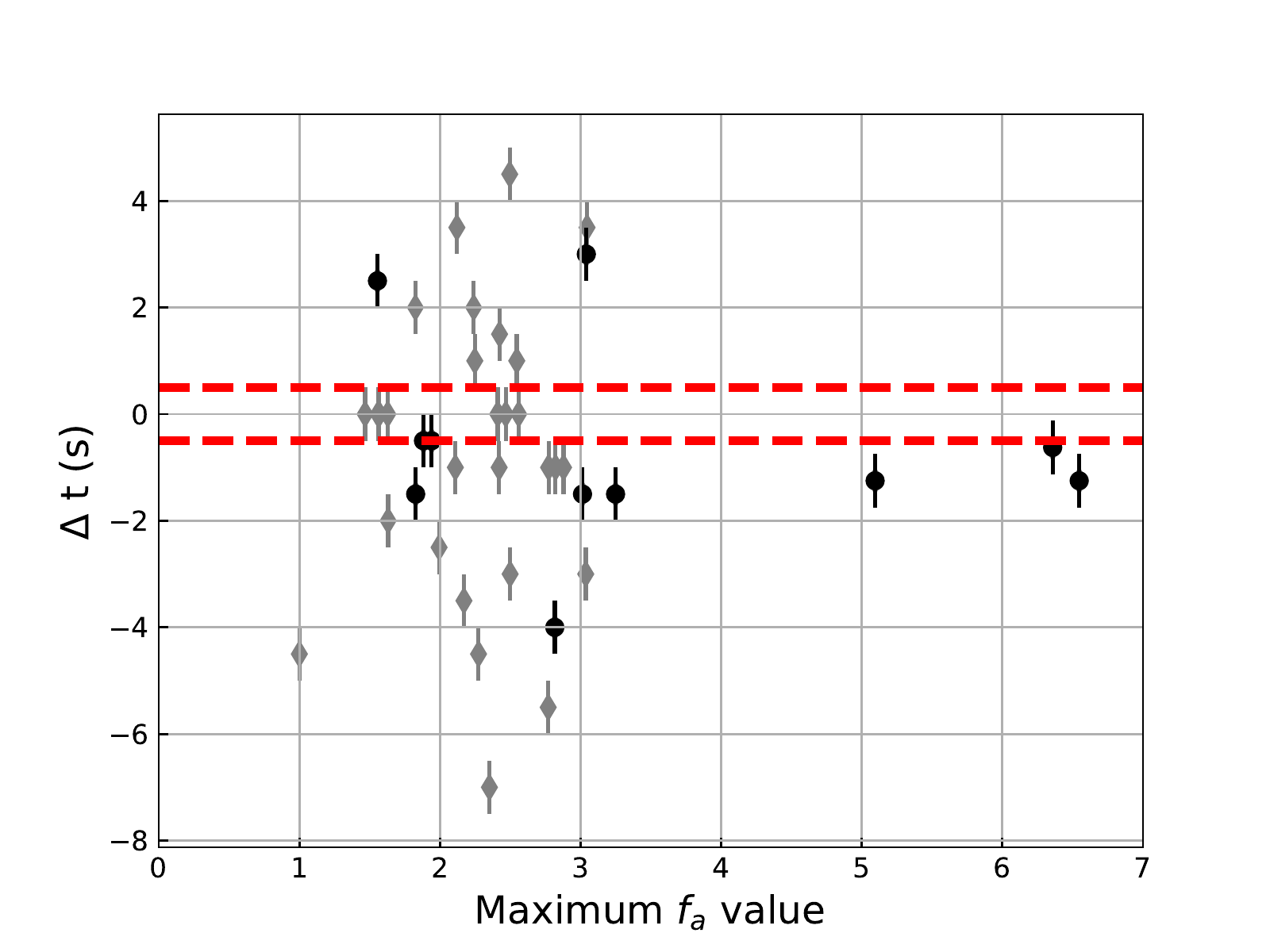}
    \caption{Time difference between the maximum \fa moment and the peak flux of a burst. In grey we show all the bursts and in black we show the cases where the difference in the \fa value when it reached its maximum and at the peak flux moment is larger than 3.0$\sigma$. Horizontal red dashed lines show the 0.5~s time interval, which is the typical exposure time of X-ray spectra at around the peaks flux moments.}
    \label{fig:lag_maxf_peakfa}
\end{figure}

As commonly observed thanks to NICER's soft X-ray coverage, \citep[see e.g.,][]{2018ApJ...855L...4K,2018ApJ...856L..37K,2019ApJ...885L...1B,2020MNRAS.499..793B,2022MNRAS.510.1577G}, we detect a significant soft excess in the burst emission if the X-ray spectra are modelled assuming a constant background and a thermal radiation from the burst itself. Statistically a model that involves applying a scaling factor (\fa) to the persistent emission spectrum during the burst is the preferred model in about two thirds of all the X-ray spectra we extracted from \source. For \source we find that the largest \fa factors are observed at around the peak fluxes of the bursts and seem to be correlated with the peak flux of the burst. This is similar to the behavior we found for the bursts from Aql~X-1 \citep{2022MNRAS.510.1577G}.
Although the \fa factor itself is simply a scaling factor to the persistent emission of the system during the burst, it has been attributed to the increased mass accretion rate during the bursts due to PR drag caused by the burst photons. Simulations of the interaction of bursts with their environment \citep{2018ApJ...867L..28F,2020NatAs...4..541F} show that indeed bursts do cause PR drag which results in the accretion rate to increase by several factors. However this increase in mass accretion rate is not sustainable, since the inner part of the disk is expected to be drained faster than it is refilled by the outer parts and therefore the mass accretion rate can decrease back even before the burst peaks. Therefore a time difference between the maximum \fa value obtained at a burst and the peak flux moment can be expected. In \autoref{fig:lag_maxf_peakfa} we show this time difference. In grey we show the cases for all the bursts and in black we show the cases where the \fa value at the peak is 3$\sigma$ different than the maximum \fa value. In general we see that there is a tendency for the time difference to be negative. Specifically, in 14 bursts \fa peaks before the burst reaches its peak flux. While these results support the PR drag scenario there are bursts where the opposite trend can also be seen.

\begin{figure}
    \centering
    \includegraphics[scale=0.5]{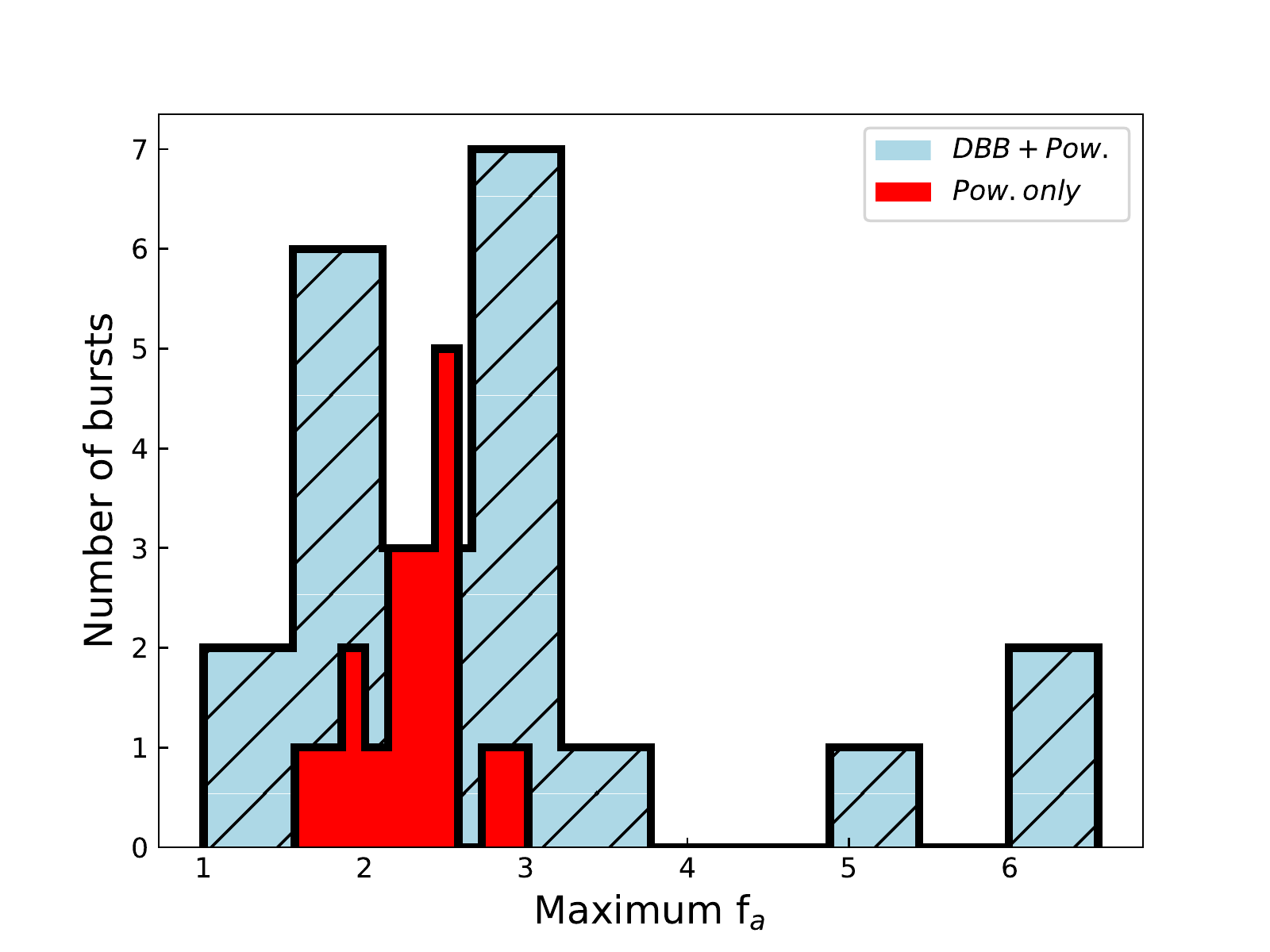}
    \caption{Maximum \fa values reached during the bursts where the addition of a disk blackbody component is required or not.}
    \label{fig:fa_rate_hist_comp}
\end{figure}

Since the \fa is related to the mass accretion rate in the inner disk, it may be expected that the maximum \fa value obtained during a burst may be related to the persistent emission of the source. Similar to our result for Aql~X-1, we found no evidence for a correlation with the persistent flux and the \fa parameter. We also checked whether the maximum \fa value can be related to the source spectral state. For this purpose in \autoref{fig:fa_rate_hist_comp} we show histograms of the maximum \fa values found for each burst in cases where a disk blackbody component is significantly detected or not. We find that the maximum \fa values attained during these bursts do not show a significant variation with each other. The only difference is that in the bursts where a disk blackbody is required the scatter of \fa values is significantly larger than the cases where the disk blackbody is statistically not needed. 

We also applied a reflection model to all the X-ray spectra where a scaling factor improved the fit. We took into account the relativistic effects on the reflection spectrum although we note that not taking into account these effects did not significantly worsen the fits. The data does not require relativistic blurring of the reflection spectra. We also tried to allow for the ionization parameters to vary through the bursts as it is expected to increase during the rise \citep{2022MNRAS.509.1736S}. However the current data often does not allow us to constrain that parameter and when allowed to vary we see that the disk is strongly ionized reaching the hard limits of the current models. 
At later stages of the bursts, the ionization parameter can be limited to lower values than the hard limits, but the results are not statistically significant.

We saw that similar to the results obtained for Aql~X-1, application of a reflection model improves the fits compared to constant background approach and the resulting statistics are comparable to results of the \fa approach. We see that at around the peak flux moments of the bursts which show photospheric radius expansion, the reflection models do not provide adequate fits but in bursts where no radius expansion is observed the reflection fraction compared to the burst emission seems to be at around 22\% level.

One similar effect of the burst emission on the system would be to cool the corona via Compton scattering of soft burst photons with energetic electrons \citep[see e.g.,][]{2003A&A...399.1151M,2018SSRv..214...15D,2020A&A...634A..58S,2020MNRAS.499.4479S}. Such a Compton cooling can be observed by detecting a decline in the hard X-ray lightcurve of the source once a burst starts. Evidence for such Compton cooling of the corona has been presented in \cite{2003A&A...399.1151M,2010A&A...510A..81C,2017A&A...599A..89K,2012ApJ...752L..34C,2013ApJ...777L...9C,2014ApJ...791L..39J,2014A&A...564A..20J,2018ApJ...864L..30C}. By stacking the \nustar~data from 7 bursts with very similar profiles, we here show further evidence that indeed the hard X-ray count rates show a decline as the burst starts and furthermore shows a negative correlation with the softer energy photons. We note however, that the hard X-ray count rates are very low and therefore the results are not definitive.

In all our analysis we tried to fit the X-ray spectra with either the \fa method or the reflection models. However, in a realistic scenario the effect of an X-ray burst on its environment is expected to cause both a detectable reflection component and the increased accretion rate. Using burst 6 as our brightest burst and 27, which is the brightest burst observed simultaneously with \nicer~and AstroSat, we tried to fit the X-ray spectra by taking into account both the reflection and a change in the persistent emission of the source (\fa method). However we see that the soft excess emission can be easily modeled with scaling the persistent emission leaving statistically no space for an additional reflection model. Once a spectrum is fitted with \fa model adding a further reflection model does not improve the fit. Furthermore, even if we force the existence of a reflection component, the flux of this component decreases by approximately two orders of magnitude lower compared to just fitting with the reflection component and becomes almost completely unconstrained. These results suggest that even larger effective area and broadband coverage will be required to be able to detect signatures of both processes at the same time in time resolved spectra of bursts, which either demands further simultaneous observations of especially brighter bursts with \nicer~ and AstroSat or larger effective area broadband observatories like STROBE-X \citep{2019arXiv190303035R} and eXTP \citep{2019SCPMA..6229502Z}.

\section*{ACKNOWLEDGEMENTS}
We thank the referee for her/his very valuable and constructive comments.
T.G. has been supported in part by the Scientific and Technological Research Council (T\"UB\.ITAK) 119F082, and the Turkish Republic, Presidency of Strategy and Budget project, 2016K121370. This work was supported by NASA through the NICER mission and the Astrophysics Explorers Program.
P.B. acknowledges support from the CRESST II cooperative agreement (80GSFC21M0002).
R.M.L. acknowledges the support of NASA through Hubble Fellowship Program grant HST-HF2-51440.001.
S.G. acknowledges the support of the CNES.
This research has made use software provided by the High Energy Astrophysics Science Archive Research Center (HEASARC), which is a service of the Astrophysics Science Division at NASA/GSFC and the High Energy Astrophysics Division of the Smithsonian Astrophysical Observatory. This paper includes data publicly available from the {\em AstroSat} mission of the Indian Space Research Organisation (ISRO), archived at the Indian Space Science Data Centre (ISSDC). 

\facilities{NICER}
\software{HEASoft, XSPEC, Sherpa, CIAO, Astropy}

\bibliographystyle{aasjournal}
\bibliography{main_text_rev1}

\appendix
\restartappendixnumbering
\section{Short Recurrence Bursts}
\label{appendix_section}

\begin{figure}
    \centering
    \includegraphics[scale=0.35]{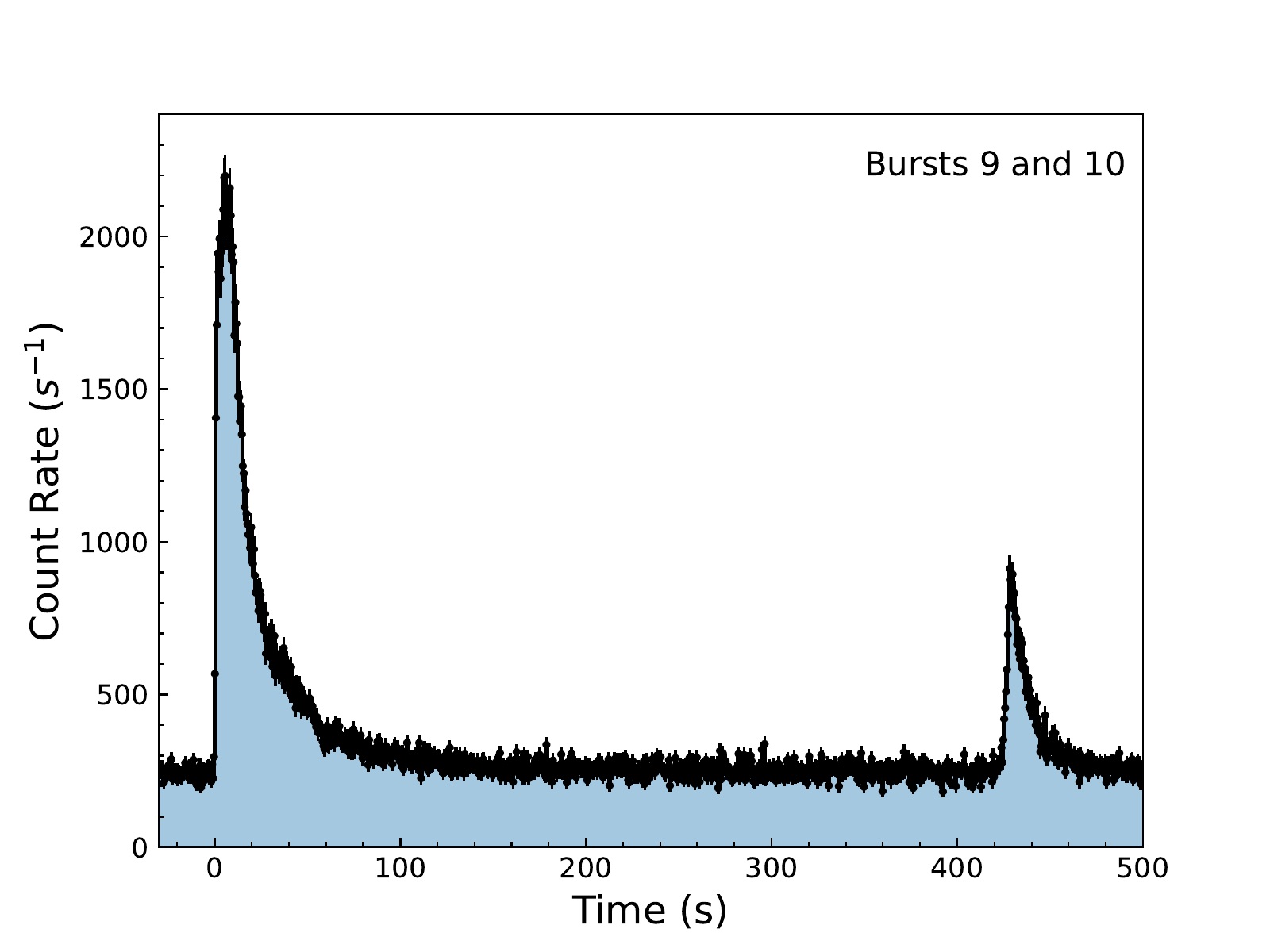}
    \includegraphics[scale=0.35]{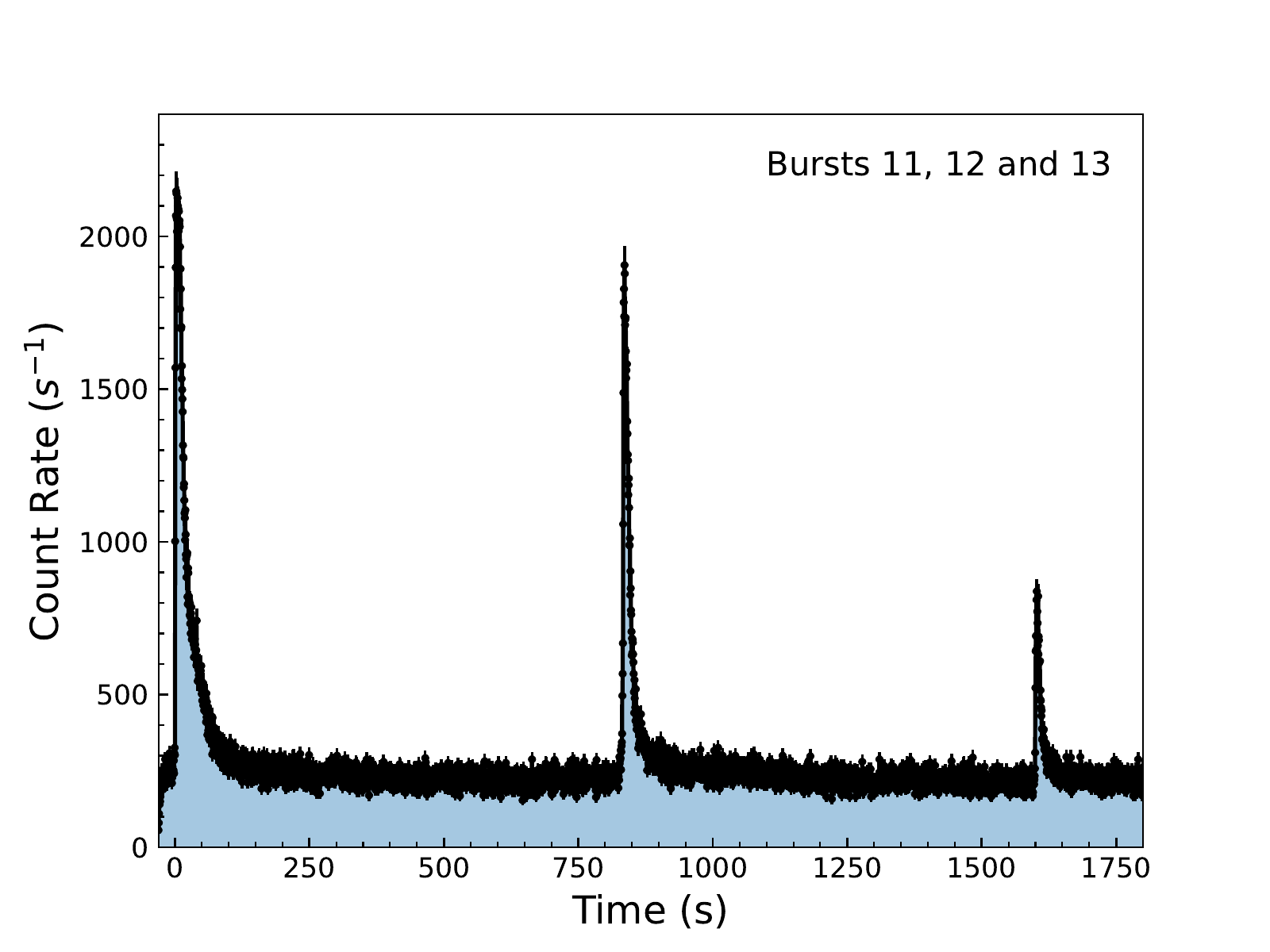}
        \includegraphics[scale=0.35]{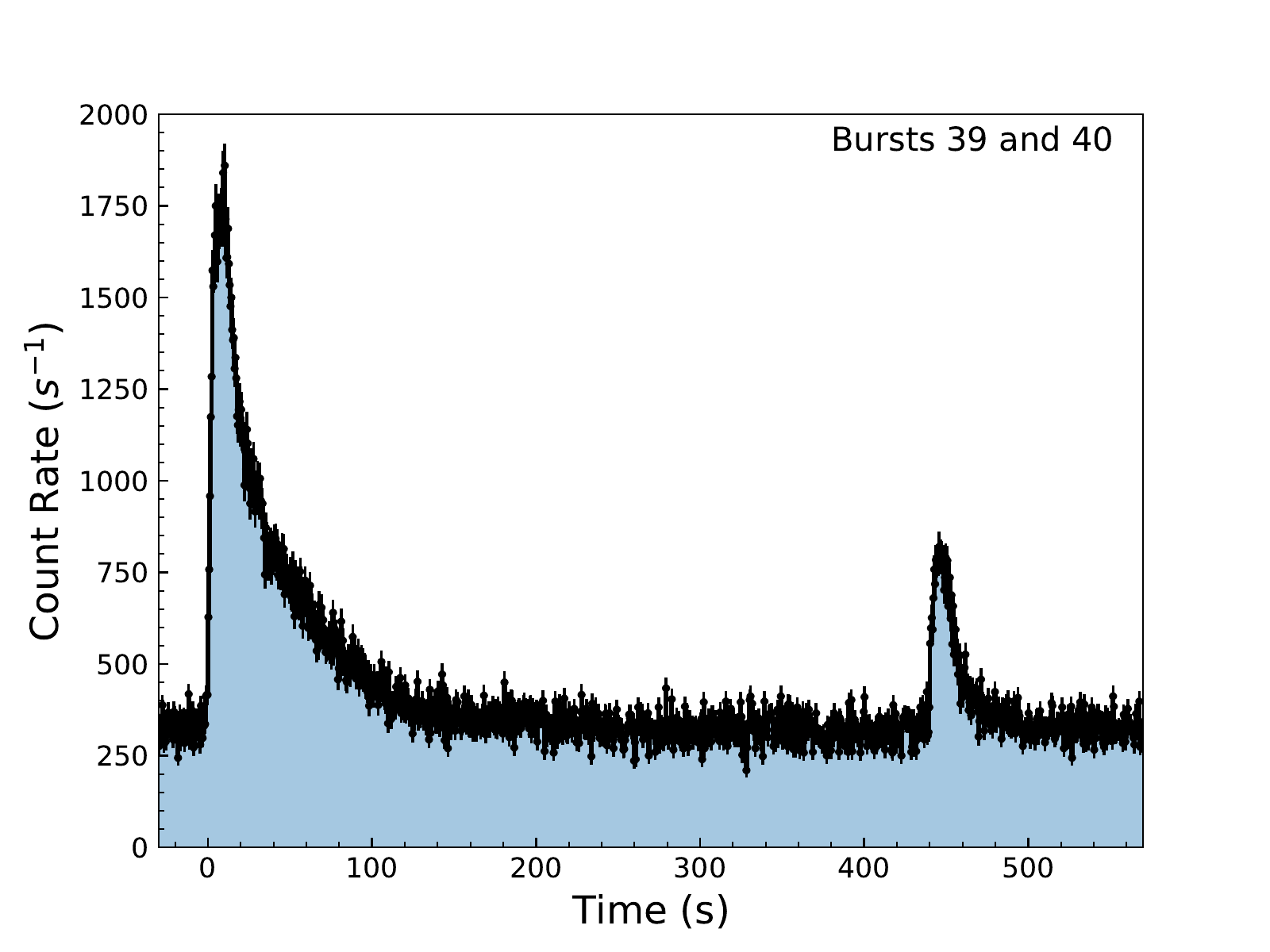}
    \includegraphics[scale=0.35]{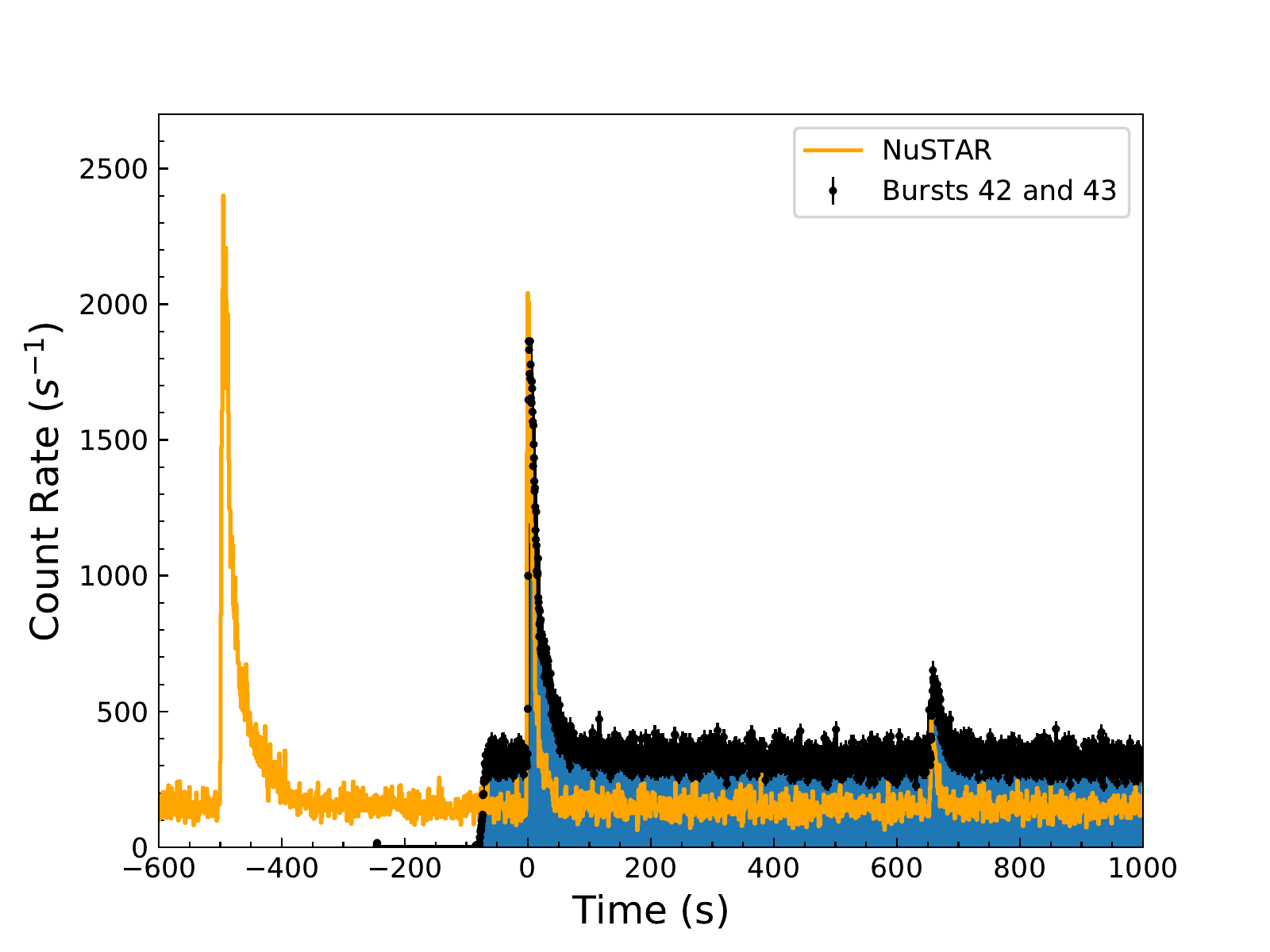}
    \includegraphics[scale=0.35]{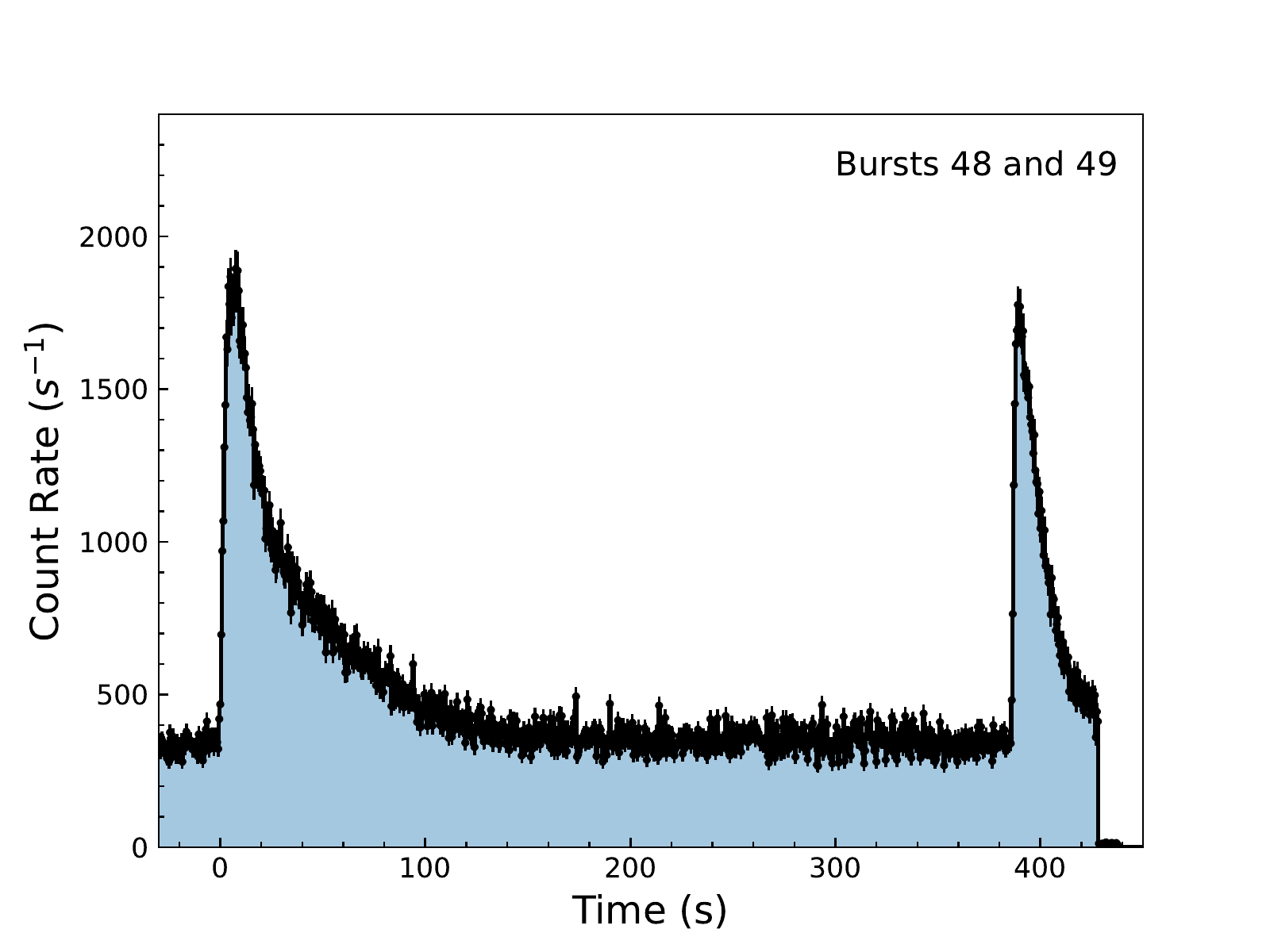}
    \caption{Lightcurves of bursts that show short recurrence. Lower-right panel also shows the overlapping \nustar~ observation where the triple nature of the bursts can be seen. For clarity 5--30~keV \nustar~ count rates are multiplied by a factor of 5.}
    \label{fig:opl_bursts}
\end{figure}

In \autoref{fig:opl_bursts} we show lightcurves of short recurrence events detected with \nicer. Bursts 9 and 10 are only separated by 425~s. Bursts 11, 12, and 13 form a triple burst event where the separations are only, 831 and 768~s between bursts 12 and 11 and bursts 13 and 12, respectively. These three bursts are separated by roughly a day with bursts 9 and 10, which is also another short recurrence event. Bursts 39 and 40 are only separated by 440~s. Bursts 42 and 43, which happen almost half a day after burst 41, are only separated by 657~s. Note that, this double burst event is actually a triple burst. Thanks to a simultaneous \nustar~ observation, we are able to identify the primary burst of this event happening roughly 450~s before burst 42 in the \nicer~ sample, which was also shown by \cite{2022A&A...660A..31Z}. Finally, bursts 48 and 49 show a very short recurrence time with a separation of only 385~s. These bursts are also worth noting because the peak count rate of the secondary burst is very comparable to the primary burst. Overall, our sample includes two triple burst events, one of which have already been reported by \cite{2022arXiv220207379R} and three double bursts. Such short recurrence bursts from \source have already been observed in various forms \citep{1982ApJ...258..254O, 1982ApJ...263..325P, 2010ApJ...718..292K, 2019MNRAS.482.4397B, 2022arXiv220207379R}.


\label{app:oplbursts}

\end{document}